# Vapor Phase Growth and Grain Boundary Structure of Molybdenum Disulfide Atomic Layers

Sina Najmaei[1†], Zheng Liu[1†], Wu Zhou[2,3†], Xiaolong Zou[1], Gang Shi[1], Sidong Lei[1], Boris I. Yakobson[1], Juan-Carlos Idrobo[3], Pulickel M. Ajayan[1], Jun Lou[1]

[1]Department of Mechanical Engineering & Materials Science, Rice University, Houston, Texas 77005, US

[2] Department of Physics and Astronomy, Vanderbilt University, Nashville, TN 37235, USA.

[3] Materials Science and Technology Division, Oak Ridge National Laboratory, Oak Ridge, TN 37831, USA.

[†] These authors contributed equally to this work
*Correspondence to:  jlou@rice.edu

**Single layered molybdenum disulfide with a direct bandgap is a promising two-dimensional material that goes beyond graphene for next generation nanoelectronics. Here, we report the controlled vapor phase synthesis of molybdenum disulfide atomic layers and elucidate a fundamental mechanism for the nucleation, growth, and grain boundary formation in its crystalline monolayers. Furthermore, a nucleation-controlled strategy is established to systematically promote the formation of large-area single- and few-layered films. The atomic structure and morphology of the grains and their boundaries in the polycrystalline molybdenum disulfide atomic layers are examined and first-principles calculations are applied to investigate their energy landscape. The electrical properties of the atomic layers are examined and the role of grain boundaries is evaluated. The uniformity in thickness, large grain sizes, and excellent electrical performance of these materials signify the high quality and scalable synthesis of the molybdenum disulfide atomic layers.**

Atomic layered graphene has shown many fascinating properties as a supplement to silicon-based semiconductor technologies [1-4]. Consequently, great effort has been devoted to the development and understanding of its synthetic processes [5-8]. However, graphene with its high leaking current, due to its zero bandgap energy, is not suitable for many applications in electronics and optics [9, 10]. Recent developments in two different classes of materials – transition metal oxides and sulfides – have shown many promises to fill the existing gaps [10-12]. For example, the successful demonstration of molybdenum disulfide ($MoS_2$)-based field-effect transistors (FET) [11], has prompted an intense exploration of the physical properties of few-layered $MoS_2$ films [13-17].

$MoS_2$ is a layered semiconductor with a bandgap in the range of 1.2-1.8 eV, whose physical properties are significantly thickness-dependent [13, 14]. For instance, a considerable enhancement in the photoluminescence of $MoS_2$ has been observed as the thickness of the material decreases [14]. The lack of inversion symmetry in single-layer $MoS_2$ and developments in controlling the available valley quantum states in this material have brought valleytronic applications closer to reality [15]. However, the large-scale synthesis of high quality $MoS_2$ atomic layers is still a challenge.

Recent top-down approaches to obtain large areas of $MoS_2$ thin films have seized considerable attention [18, 19]. Nevertheless, the lack of uniformity in thickness and size undermines the viability of such approaches. Other techniques for the scalable synthesis of $MoS_2$ have focused on the direct sulfurization of molybdenum-containing films. Recent work on the solid phase sulfurization of molybdenum thin films revealed a straightforward method for large-area synthesis of $MoS_2$ [20]. However, the suboptimal quality of samples prepared using this synthetic route – exemplified in their low carrier mobility (0.004 - 0.04 $cm^{-2}V^{-1}s^{-1}$), as compared to mechanically exfoliated samples (0.1 - 10 $cm^2V^{-1}s^{-1}$), and small grain sizes (~20 nm) –remains a problem [20]. Alternatively, Liu *et al.* have exploited the sulfurization of ammonium tetrathiomolybdate $(NH_4)_2MoS_4$ films [21]. Yet, this method bears a considerable level of complexity in precursor preparation and in achieving sufficient quality, which has limited its feasibility [21].

Traditionally, the sulfurization of molybdenum trioxide ($MoO_3$) has been the primary approach in $MoS_2$ nanomaterial synthesis and it has been widely studied [22-28]. Making use of its low melting and evaporation temperatures, Lee *et al.* demonstrated that $MoO_3$ is a suitable precursor for chemical vapor deposition (CVD) of $MoS_2$ thin films [22]. This work studied the nucleation and growth process in CVD grown $MoS_2$ atomic layers facilitated by seeding the substrate with graphene like species, but lacked a comprehensive characterization of grains and grain boundary structures[22]. The evolution in graphene synthesis suggests that an in-depth understanding of the CVD process for single-crystalline domain nucleation and growth is essential to the large area preparation of high quality materials [5-8]. Here, we develop a CVD-based procedure for the large-area synthesis of highly-crystalline $MoS_2$ atomic layers by vapor-phase $MoO_3$ sulfurization. Furthermore, we evaluate the growth process, grain morphology, and grain boundary structure of the polycrystalline $MoS_2$ atomic layers and characterize their corresponding electrical performances.

To synthesize $MoS_2$ atomic layers in a vapor phase deposition process, $MoO_3$ and pure sulfur were applied as precursor and reactant materials, respectively (supplementary Fig. S1). To understand the growth process of $MoS_2$, the source supply was controlled by dispersing $MoO_3$ nanoribbons with varied concentrations on a substrate, away from the designate growth region, while all other parameters were kept constant. Insulating $SiO_2$/Si substrates were used, and the experiments were performed at a reaction temperature of 850$^o$C. The experiments show that the synthesis of $MoS_2$ was limited by the diffusion of vapor phase $MoO_{3-x}$: varying the density of dispersed nanoribbons systematically slows or stops the growth at certain points, providing means to explore the growth process. Several stages were observed during the $MoS_2$ atomic layer growth. Initially, small triangular domains were nucleated at random locations on the bare substrate (Fig. 1a). Then, the nucleation sites continued to grow and formed boundaries when two or more domains met (Figs. 1b and 1c), resulting in a partially continuous film. This process can eventually extend into large-area single-layered $MoS_2$ continuous films if sufficient precursor supply and denser nucleation sites are provided (Fig. 1d). Through additional experiments, we determined that sulfur concentration and chamber pressure are the two key parameters that control the prospects of the material growth (Supplementary Figs. S2 and S3). These parameters and their synergistic effects on the morphology, nucleation and growth of MoS$_2$ crystals were used to control the MoS$_2$ synthesis in our experiments as detailed in the supplementary information. These observations show that, in contrast to graphene that nucleates abundantly on copper substrates without any treatment [8], the limiting factor in the growth of large-area highly-crystalline MoS$_2$ films is the rare and complicated nucleation process of this material on bare SiO$_2$ substrates. These observations are in agreement with previous studies on CVD growth of MoS$_2$ [22] and elucidate the importance of efficient control on nucleation, essential to the large area growth of the MoS$_2$ atomic layers.

In the quest for feasible strategies to control the nucleation process, we take advantage of some of our common experimental observations. Our experiments show that the MoS$_2$ triangular domains and films are commonly nucleated and formed in the vicinity of the substrates' edges, scratches, dust particles, or rough surfaces (supplementary Fig. S4). We utilized this phenomenon to control the nucleation by strategically creating step edges on substrates using conventional lithography processes (Fig. 1e). The patterned substrates with uniform distribution of rectangular SiO$_2$ pillars (40×40 μm$^2$ in size, 40 μm apart, and ~40 nm thick) were directly used in the CVD process for MoS$_2$ growth (Fig. 1e). The pillars facilitate a high density of domain nucleation and the continued growth allows for the formation of large area continuous films (supplementary Fig. S5). Raman spectroscopy measurements indicate that the MoS$_2$ atomic layers grow on the surface of the pillars as well as on the valley space between them, suggesting that both the surface roughness and edges may contribute to the nucleation process (Figs. 1f, and 1g). The as-grown films are predominantly single-layered, with small areas consisting of two or few layers at the preferred nucleation sites (Fig. 1g). Our observations show that the pattern base growth process follows pressure and sulfur concentration dependencies similar to the growth of MoS$_2$ on pristine substrates. The strategies to enhance this approach, such as variations in the size and geometry of the pillars and their effects on the growth process are discussed in the Supplementary Materials (supplementary Figs. S6, and S7). The inherent dependence of this approach on the edge-based nucleation resembles some of the observations and theoretical predictions in the growth other layered materials [29-30].

Theoretical studies have revealed a significant reduction in the energy barrier of graphene nucleation close to the step edges, as compared to flat surfaces of transition metal substrates [30]. We propose that similar edge-based catalytic processes are also involved in the initiation of systematic $MoS_2$ growth. The simple lithography methods applied in the preparation of silicon oxide patterns and step edges on insulating substrates provides a unique and robust strategy for the growth of large-area high-quality $MoS_2$ films that can be readily transferred to any substrate or applied to fabricate devices (Fig. 1h). This strategy can potentially be adopted to grow large area $MoS_2$ films on other insulating substrate such as hexagonal boron nitride, to take advantage of its atomically smooth surface and possible electrical performance enhancement. Further experiments reveal that growth process in these films also resembles the process described for nucleation and growth on bare substrates; they start by nucleation of triangular domains, followed by their continued growth and coalescence. As we establish the knowledge of the $MoS_2$ atomic layer growth, it becomes evident that the nucleation and growth of the triangular domains, as well as the formation of boundaries among these domains, play an important role in the development of large-area $MoS_2$ atomic layers. Therefore, the characterization of the triangular domains becomes essential to understanding the advantages and limitations of this synthetic route and the proposed growth mechanism.

A representative atomic force microscopy (AFM) image of a $MoS_2$ triangular domain is shown in Fig. 2a. The thickness of these triangles is approximately 0.7 nm, corresponding to one $MoS_2$ atomic layer. The uniformity in thickness of these samples was further confirmed by Raman spectroscopy (supplementary Fig. S8). The chemical composition of the sample was confirmed by X-ray photoelectron spectroscopy (XPS) and Electron Energy Loss Spectroscopy (EELS) analysis (supplementary Fig. S9). The equilateral geometry of the triangles with perfect $60°$ angles suggests the single-crystal nature of these nanostructures, with edges parallel to a specific lattice orientation (Fig. 2a). Transmission electron microscopy (TEM) and aberration-corrected scanning transmission electron microscopy (STEM) were then applied to examine the crystal structure of $MoS_2$ triangles and atomic layers. The selected area electron diffraction (SAED) pattern from a typical $MoS_2$ triangle (Fig. 2b) presents only one set of six-fold symmetry diffraction spots, confirming that these triangles are single-crystalline with hexagonal $MoS_2$ structure. High-resolution TEM (HRTEM) images from the edge of the triangles (Fig. 2c) clearly show fringes with ~0.6 nm spacing, corresponding to single-layered $MoS_2$. The atomic structure of the single-layer triangles is shown by the STEM annular dark field (ADF) image in Fig. 2d, where the alternating brighter and dimmer atomic positions in the hexagonal rings correspond to Mo and $S_2$, respectively. Quantitative analysis of the STEM-ADF image intensity confirms that the sample is single layer $MoS_2$ with two S atoms at each $S_2$ site [31]. Moreover, a uniform and almost defect-free structure was observed across the entire single-crystal $MoS_2$ domain. Nevertheless, through formation of grain boundaries when these single-crystalline domains merge, polycrystalline films of $MoS_2$ form. It is known that the grain boundaries in polycrystalline two dimensional materials degrade their physical properties and to overcome these problems large grain sizes are desirable[32]. Therefore characterization of crystallinity in the large area $MoS_2$ films is necessary to assess the quality of the material.

We evaluate the crystallinity of the large-area continuous $MoS_2$ films by examining the microstructure of these atomic layers using dark-field (DF) transmission electron microscopy (TEM) imaging technique [33-35]. A total of 507 DF TEM images acquired from the same diffraction spot were used to construct the DF TEM image (supplementary Fig. S10). The ~15×15 µm region shown in Figure 3a contains two areas of single-layer $MoS_2$ and demonstrates a contrast between the different areas of uniform crystal orientations. The white regions represent single-layered $MoS_2$ with the same lattice orientation. The DF TEM images, along with additional information presented in the supplementary work, demonstrate the large grain sizes in the $MoS_2$ films (several tens of microns). This is in agreement with the average size of triangular domains observed in our experiments before they coalesce and reaffirms our hypothesis for $MoS_2$ growth process. Additionally, the thickness uniformity over regions greater than tens of microns further exemplifies the high quality of these samples. Nevertheless, grain boundaries that mediate the crystal orientation transitions between the grains are one of the major sources of defects in materials that deserve closer scrutiny.

An example of grain boundary in $MoS_2$ atomic layers is shown in Figure 3b, where the selected area electron diffraction (SAED) pattern collected from the highlighted area reveals three sets of six-fold symmetry spots (Fig. 3c), corresponding to the relative rotation of ~5° and 30° in crystal orientations among the three grains. The location of the grains and their boundaries can be identified from the false-colored DF TEM image presented in Figure 3d. The image exemplifies two possible mechanisms of boundary formation observed in $MoS_2$. We hypothesize that in some occasions, traditional grain boundaries consisting of chemical bonding between the two single-layered grains are formed and the in-plane growth stops; at times the growth continues by the nucleation of a second layer at the boundaries (supplementary Fig. S11a), as seen in the marked pale blue stripe along the grain boundary between the green and blue grains shown in Fig. 3d. This is a common feature even visible in optical images that can be used to distinguish the grain boundaries. The other mode of boundary formation occurs when one grain continues to grow on top of the other without forming chemical bonding between the two grains, resulting in a bi-layered region (supplementary Fig. S11b). The mechanisms of grain boundary formation can be explored by taking a closer look at the structure of the grain junctions in $MoS_2$ atomic layers.

A combination of aberration-corrected STEM ADF imaging and first-principles calculations based on density functional theory (DFT) was applied in order to provide an in-depth analysis of the atomic configuration at grain boundaries in single layers of $MoS_2$. It has been well documented that grain boundaries in graphene can be described as arrays of dislocations formed by 5|7 fold carbon rings [33-35]. However, little is known about the detailed grain boundary structures in binary systems such as h-BN and $MoS_2$ [36-37]. Identification of the detailed atomic structure of dislocations and grain boundaries in h-BN and $MoS_2$ has been primarily a theoretical investigation, due to the lack of high-quality samples. Figure 4a shows an atomically-resolved STEM-ADF image from a grain boundary with ~21° tilt angle in false colors. The darker atomic columns in Fig. 4a correspond to two S atoms on top of each other, while the brighter spots are single Mo atoms. A detailed analysis of the STEM-ADF image indicates that all dislocations at this observed grain boundary have the shortest Burgers vector (1, 0), following the same notation for Burgers vector in graphene [32] (supplementary Fig. S12), strongly suggesting that these dislocations can be constructed by removing or inserting a semi-infinite stripe of atoms along the armchair direction. A closer look at the regions highlighted by the light blue and red rectangles in Fig. 4a reveals the atomic configurations of dislocation cores that are commonly observed at the $MoS_2$ grain boundaries. The Mo-oriented dislocations consist of 5|7 fold rings (Fig. 4b) and their variations, with sulfur substitutions at one of the sharing Mo sites (Fig. 4c).

Basic dislocation theory [38], combined with DFT calculations were employed to understand the atomic structure and formation energy of the pristine and sulfur-substituted 5|7 fold rings (Supplementary information). Due to the unique triple-layered structure of $MoS_2$, the dislocation cores extend three-dimensionally, with two layers of five-fold or seven-fold rings joining at Mo sites in the middle layer (Figs. 4d, 4e, 4f). The rings are comprised of a Mo-oriented dislocation core structure formed by fifteen atoms, with two Mo atoms constituting a Mo-Mo bond shared by the 5|7 fold rings (Fig. 4e). In each Mo-oriented 5|7 dislocation, there are five distinct Mo sites that could be substituted by S atoms (Fig. 4e). At Mo sites 1, 4, and 5, six S-S homo-elemental bonds are introduced after $S_2$ substitution (*i.e.* two S atoms replacing one Mo), while the sharing Mo-Mo bond in the middle remains unchanged. In contrast, for $S_2$ substitution at the middle Mo sites 2 and 3, the Mo-Mo bond is replaced by two Mo-S bonds, and only four S-S bonds are generated, making this type of $S_2$ substitution energetically favorable (Fig. 4f). In sulfur-rich conditions, the formation energies for the Mo-$S_2$ substitution at sites 1-5 are 5.2, -0.8, -0.9, 1.0, and 1.5 eV, respectively. With the lowest negative formation energy, the $S_2$ substitution of Mo site 3 is the most stable among the various possible structures, consistent with our experimental observations (Fig. 4c). In addition to the aforementioned traditional grain boundaries, overlapped junctions may also form when two $MoS_2$ grains merge.

Figure 4g shows a false-color DF TEM image from two grains with 30° rotation. As a result of the formation of in-plane covalent bonds, the red and green grains form grain boundaries in the upper part, while in the lower part of the junction, the two grains continue to grow on top of each other, forming a bi-layered region at the junction. A closer look at this overlapped junction, Fig. 4h, exposes the distinct Moiré pattern which combined with the two sets of six-fold symmetry diffraction spots (shown in the inset), corroborate the claim that only two grains are present in this region. By Fourier filtering each set of diffraction spots, the spatial extension of the grains can be mapped out (Fig. 4i and supplementary Fig. S13). The absence of discontinuity in the lattice structure within each grain at the point of junction suggests that the two layers continue to grow on top of each other without forming any in-plane chemical bonds. Consequently, the interaction between these grains at the overlapped junctions could be mediated by weak Van der Waals forces. Moreover, Fig. 4g demonstrates that the formation of conventional grain boundaries and the overlap of junctions are competing processes, but a clear dependence on the degree of lattice orientation mismatch in the two grains is not observed in our experiments. The existence of either type of these grain boundaries is not desirable for high-performance $MoS_2$ based electronic devices. Additionally, since transport in $MoS_2$ is highly anisotropic and the in-plain conduction is more favorable [39], the impeding effects of over-layered grain boundaries should be more dire, and examination of the CVD grown $MoS_2$ samples with the combination of these grain boundaries is imperative.

To evaluate the electrical performance of the materials, we measure their field effect carrier mobilities and compare with measurements on the exfoliated samples reported in the literature. FET devices were fabricated by patterning the films with the commonly-used lithography techniques and reactive ion etching (Fig. S15). These devices have a channel length and width of 100 μm and 10 μm, respectively. All devices demonstrated FET characteristics of n-type semiconductors, similar to the measurements on mechanically-exfoliated samples (Fig 5a) [11]. We estimate the charge carrier mobility in these devices using the equation $\mu = [dI_{ds}/dV_{bg}] \times [(L/(WC_iV_{ds}))]$, where L and W are the channel length and width. The capacitance between the channel and the back gate per unit area is estimated to be ~$1.2 \times 10^{-4}$ Fm$^{-2}$ ($C_i = \varepsilon_0\varepsilon_r/d$, where $\varepsilon_0$=3.9 and d=285 nm). The theoretical estimate for the capacitance of $SiO_2$ will provide us with a lower limit estimate for the calculated mobilities. Mobility measurements obtained under good ohmic contact (Fig. 5a and inset) show an average of 4.3 ± 0.8 cm$^2$V$^{-1}$s$^{-1}$ for the multiple devices with thicknesses ranging from single to multiple layers (Fig. 5b), which is within the range of reported experimental values for mechanically-exfoliated samples. The measured ON/OFF ratios for the devices reach a maximum value of $6\times10^6$ for the gate voltage in the range of -150 to 150 V and a source-drain bias of 5 V, comparable to measurements done by Radisavljevic *et al.* [11]. We attribute the small differences in the electrical properties of our samples, as compared to the natural crystals, to the expected higher density of defects in the synthesized materials, which degrades the gating.

It is known that grain boundaries play an important role as scattering centers that degrade the charge carrier mobility in the materials. Our calculations on the static electronic structures of various grain boundaries in $MoS_2$ – provided in the supplementary information (supplementary Fig. S16) – show the presence of local states deep inside the gap at Mo sites along the grain boundaries, which undermine the electronic transport of $MoS_2$. To quantify the effects of grain boundaries on the electrical performance of CVD grown $MoS_2$, we perform channel length dependent electrical measurements. Our analysis of polycrystalline $MoS_2$ shows that the grain boundaries are a preferred nucleation site for the growth of secondary layers (Fig. S17). This common characteristic is detectable optically and confirmed by dark field-TEM imaging (Fig. 5c and Fig. 5d). For these experiments, large area films were patterned into ribbons and FET transistors with different channel length were fabricated (Fig. 5e). The optical images acquired from these devices reveal the characteristic stripes along the grain boundaries across the channels at deferent locations (Fig. 5f). Raman analysis shows that these strips are bi-layered regions and provides a clear location map of the defects (Fig. 5g). It is prudent to assume that at least part of these bi-layered stripes, if not all, represent the grain boundaries in the material, and that a higher density of these strips is seen in the longer channel lengths (Figs. 5f and 5g). Mobility measurements as a function of channel length were performed using drain-source voltage of 0.1 V and in the gating range of ±40V. The mobility measurements show an average decrease by roughly 50% as the channel length increases from 3 to 80 microns (Fig. 5h). Under these conditions the device does not reach its saturation current and the on/off current ratios do not represent the maximum capacity of the devices, however these values were measured and compared as a function of channel length for the sake of evaluating grain boundary effects (Fig. 5i). The measurements show a decrease of roughly one order of magnitude in the on/off current ratios as the channel length is increased to 80 μm. It is worth noting that the work by Liu et al. on understanding the short channel effects in exfoliated $MoS_2$ details the short channel effects on the transport properties of $MoS_2$ [40]. They reveal that for channel lengths larger than ~ 1 μm, as the transport is predominantly in the diffusive regime, the mobility and on/off ratio are expected to saturate at a constant value [40]. Assuming that the impurity concentration is roughly uniform across the channel, it can be deduced that the observed changes in our experiments is stemmed from the increased density of line defects introduced by the grain boundaries (Fig. S18). These defects increase the electron scattering and decrease the mobility of the electrons and in addition the local changes in the band structure induced by these defects, degrade the gating process and reduce the on/off current ratio (Fig S16). However, the trapped substrate charges and surface interactions are potentially of higher importance in decreasing the magnitude of these transport properties. Therefore, mechanisms of high dielectric top-gating and shorter channel lengths could significantly enhance the device properties, as it suppresses the scattering of the carriers caused by surface charge traps on the substrate and the grain boundaries and enhance the gating experiments[11, 41].

In conclusion, our results demonstrate the vapor phase growth of $MoS_2$ atomic layers by nucleation, growth and grain boundary formation of single crystalline domains. A straightforward method for the controlled nucleation of molybdenum disulfide and large area synthesis of films was demonstrated. The coalescence of the grains lead to the formation of grain boundaries composed of arrays of dislocations of pristine and sulfur-substituted 5|7 fold rings. Alternatively, some grains are joined by simply growing on top of one another without forming any chemical bonds. The structure and the electrical retribution of these boundaries are assessed in depth. The high quality of these atomic layers, exemplified by their electrical performance, presents a significant step towards scalable preparation of this material.

**Methods**

A scanning electron microscope (SEM, FEI Quanta 400) is used to study the morphology and the distribution of $MoS_2$ triangles and films on silica substrates. Raman spectroscopy performed on a Renishaw inVia microscope characterizes the structure and thickness of the films at 514.5 nm laser excitation. Film thickness and topographical variations in the samples were measured using Atomic force microscopy (AFM, Agilent PicoScan 5500).

Transmission electron microscopy is employed to study the morphology, crystal structure, and defects of the $MoS_2$ samples. The TEM experiments are performed on a JEOL 2100F microscope operating at 200 kV, and a FEI Titan 80-300 TEM operating at 60 kV. The diffraction patterns of individual $MoS_2$ triangles are obtained from the whole triangle using a selected-area aperture. For the large-size DF TEM image of the $MoS_2$ film (supplementary Fig. S10), 507 DF TEM images are taken at the same diffraction spot under the same microscope settings, with spacing of 3 µm for each frame, under the DF imaging mode. The images are then aligned and stitched by Microsoft Image Composite Editor. To recheck every composite, BF TEM images at low magnification are acquired (supplementary Fig. S10a). False-color DF TEM images are obtained by Red, Green, and Blue (RGB) image construction of individual DF images acquired at the DF mode by selecting the spots at each pattern set with a small objective aperture.

Scanning transmission electron microscopy (STEM) imaging and spectroscopy analysis are performed on an aberration-corrected Nion UltraSTEM-100 operating at 60 kV [42]. The convergence semi-angle for the incident probe is 31 mrad. Annular dark-field (ADF) images are gathered for a half-angle range of ~86–200 mrad. EEL spectra are collected using a Gatan Enfina spectrometer, with an EELS collection semi-angle of 48 mrad. The ADF images presented in this manuscript are low-pass filtered in order to reduce the random noise in the images. The ADF image shown in Fig. 4a is prepared by partially filtering the direct spot in the FFT of the image. This filtering process helps remove the contrast generated by the surface contaminations. A comparison of the ADF images before and after the partial direct spot filtering is shown in the supplementary Fig S14.

TEM samples are prepared using a standard PMMA-based transfer, where the samples are spin-coated and submerged in basic solutions such as KOH. Subsequently, acetone and DI water are used to clean the TEM grid, allowing for the suspension of the films in the solution and the transfer to TEM grids. We note that this method tends to damage the $MoS_2$ film by creating small holes, as seen in Fig. 3b. Samples for STEM analysis are prepared by placing the as-grown $MoS_2$ films onto lacey-carbon TEM grids, and immersing them into 2% HF solution for 20 seconds. Most of the $MoS_2$ films and triangles are then transferred onto the grid.

X-ray photoelectron spectroscopy (XPS, PHI Quantera) is performed using monochromatic aluminum KR X-rays, and the obtained data is analyzed with the MultiPak software. Field effect transistor (FET) devices are made by photolithography process using photoresist S1813, mask aligner (SUSS Mask Aligner MJB4), and $O_2$ reactive ion etching (RIE, Phantom III). The photoresist is removed by acetone and PG-REMOVER. A similar lithography process is used to prepare the photoelectric devices, with the addition of the photoresist LOR5B as an adhesive layer to achieve an undercut. The electrodes (Au/Ti, 50nm / 3nm) are deposited using the e-beam evaporator. Electrical measurements are performed in a costume-built probe station under vacuum conditions ($10^{-5}$ Torr), and field-effect transistor (FET) and I-V measurements are carried out using two, Keithley 2400 source meters. The photoelectric measurements are performed in a home-built electrical and environmental noise-reducing chamber, using a Stanford Research SR830 Lock-in Amplifier combined with an optical chopper for signal modulation and noise filtration.

**Acknowledgements**

This work was supported by the Welch Foundation grant C-1716, the NSF grant DMR-0928297, the U.S. Army Research Office MURI grant W911NF-11-1-0362, the U.S. Office of Naval Research MURI grant N000014-09-1-1066, and the Nanoelectronics Research Corporation contract S201006. This research was also supported in part by National Science Foundation through grant No. DMR-0938330 (WZ); Oak Ridge National Laboratory's Shared Research Equipment (ShaRE) User Program (JCI), which is sponsored by the Office of Basic Energy Sciences, U.S. Department of Energy. The computations were performed at the Cyberinfrastructure for Computational Research funded by NSF under Grant CNS-0821727 and the Data Analysis and Visualization Cyberinfrastructure funded by NSF under Grant OCI-0959097

**Author contributions**

J.L., P.M.A, J.C.I, and B.I.Y proposed and supervised the project. S.N developed and performed the synthesis experiments. S.N, Z.L, G.S, and S.D performed the material and device performance characterizations. Z.L performed the DF-TEM imaging experiments. W.Z. performed the STEM characterization and part of the DF-TEM imaging experiments. X.Z performed the grain boundary calculations. All authors helped with analyzing and interpreting the data. S.N, Z.L, W.Z, X.Z, and J.L wrote the paper and all authors discussed and revised the final manuscript.

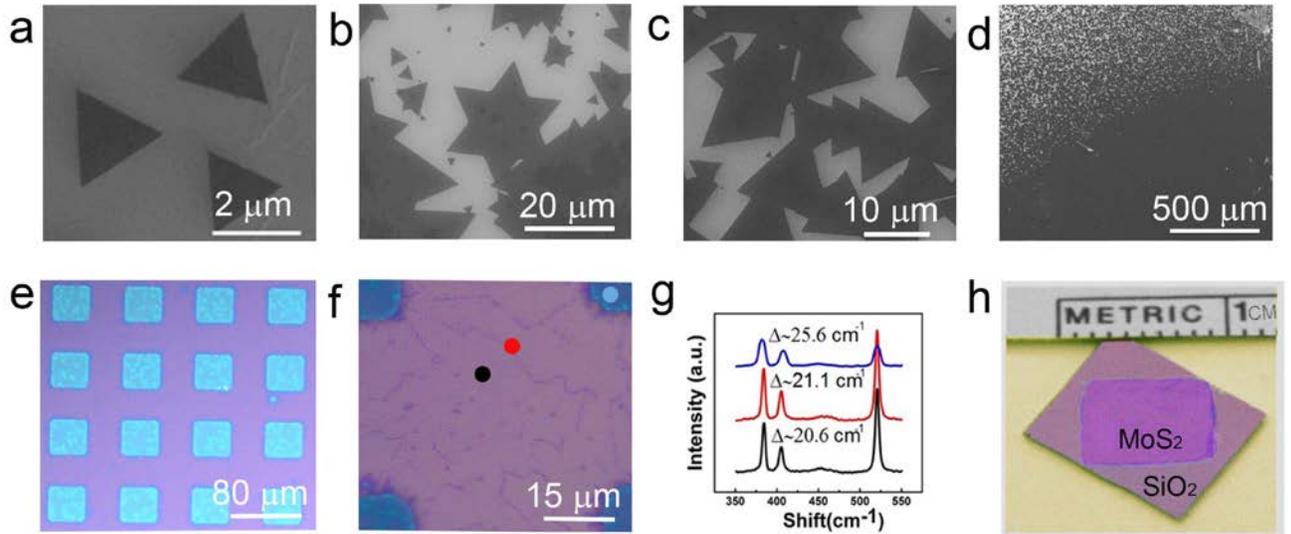

**Figure 1| The fundamental growth process of MoS₂ films and controlled nucleation. a-d,** SEM images showing the growth process of MoS$_2$ from small triangles to continuous films. **e,** Optical image of a large area continuous film synthesized on substrates with rectangular shaped patterns. These patterns are distinguishable by their bright blue color and the areas between them are completely covered by continues MoS$_2$ layers. **f,** The close up optical image of the as synthesized MoS$_2$ film, on the patterned substrate demonstrating that the method described typically results in single- and bi-layered films between the pillars and thicker samples on the pillars, these patterns act as nucleation promoters. **g,** Raman spectra acquired from different regions highlighted in Figure g, showing the thickness and its small variability in the sample. **h,** Large area, square centimeter film transferred to a new substrate from a patterned substrate using conventional polymer-base transfer techniques

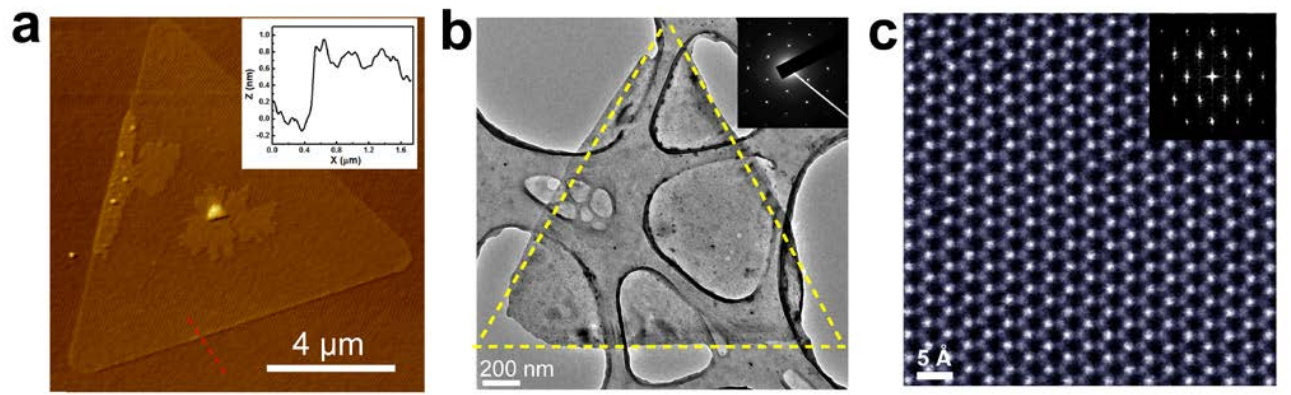

**Figure 2| Characterization of the MoS$_2$ triangular domains**. **a,** AFM image from a MoS$_2$ triangular flake. Nucleation sites at the center of the MoS$_2$ triangle are also visible. Inset: line scan demonstrating the thickness of a single layered MoS$_2$. **b,** Bright field TEM image of the MoS$_2$ triangular flakes. Inset: the SAED pattern acquired from this triangle demonstrates it is a single crystal with hexagonal structure. **c,** STEM-ADF image representing the defect free hexagonal structure of the triangular MoS$_2$ flake. The

brighter atoms are Mo and lighter ones are S2 columns. Inset: the FFT pattern of the image confirming the hexagonal MoS$_2$ structure.

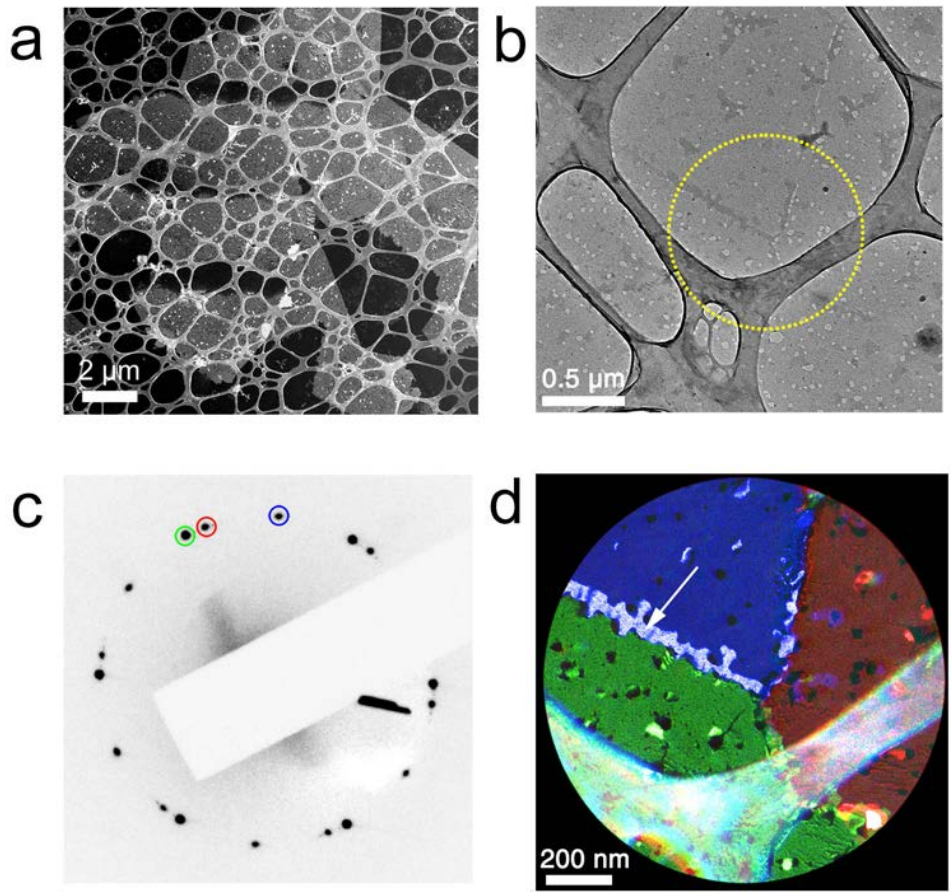

**Figure 3| Grains and Grain boundaries in MoS$_2$ films. a,** Dark field (DF) TEM image acquired from the continuous MoS$_2$ film using the same diffraction spot (collaged from ~159 individual DF images). Most of the 15x15 um imaging area is covered by grains with the same crystal orientation, demonstrates the large single-crystal grain size of the MoS$_2$ film. **b,** Bright field TEM image of the MoS$_2$ atomic layer containing a junction of three different grains. **c,** SAED pattern acquired from the highlighted area in **b** showing three sets of six-fold symmetry diffraction patterns with relative rotations of ~ 5$^o$ and 30$^o$. **d,** False-color dark field TEM image using the three diffraction spots marked in **c** showing the presence of three grains.

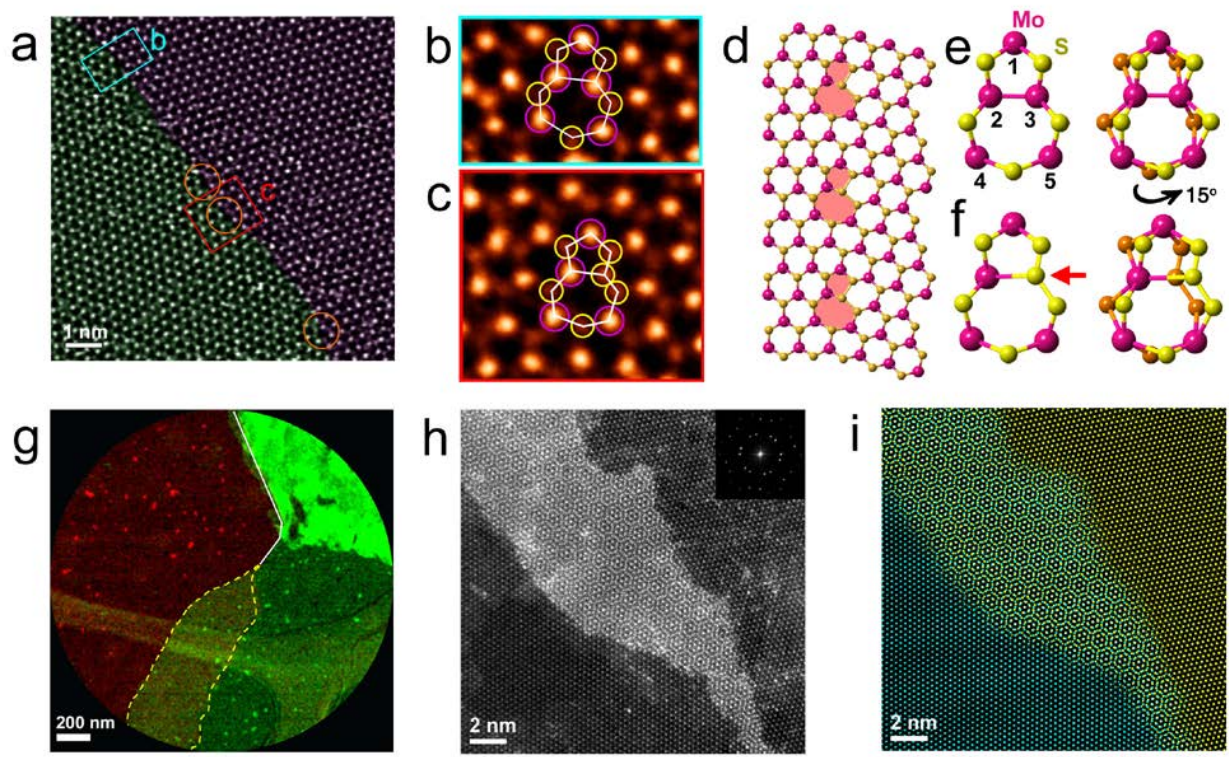

**Figure 4| Atomic configurations for grain boundaries and overlapped junctions in the single-layered MoS$_2$ films. a,** STEM-ADF image of a grain boundary with 21° tilt angle with common dislocations in MoS$_2$ highlighted. **b-c,** Magnified views of the regions highlighted by the light blue and red rectangles in **a**, respectively. Schematic of the dislocation core structures are overlaid in the images, where purple circles are Mo and yellow ones are S$_2$ columns. **d,** Structure model for a predominant form of dislocations observed in MoS$_2$ grain boundaries. **e,** Structure model for the Mo-oriented 5|7 dislocation core, experimentally confirmed and highlighted by the light blue rectangle in **a**. The numbers label different Mo sites that can be substituted by S atoms. **f,** Structure model for the Mo-oriented 5|7 dislocation core with S$_2$ substitution at the middle Mo site, experimentally confirmed and highlighted by the red circles in **a**. The yellow, orange and purple spheres represent top and bottom S and Mo, respectively. **g,** False-color dark field TEM image showing the presence of two grains with 30° rotation. The white solid line marks the grain boundaries between the two grains, while the yellow dash line marks the overlapped grain junction. **h,** STEM-ADF image of an overlapped junction between two grains. Inset: FFT of the image showing two sets of MoS$_2$ diffraction sports with 17° rotation. **i,** False-colored FFT-filtered image of **h** showing the distribution of the two grains in cyan and yellow. Overlapping of the two grains at the junction generated the distinct Moiré fringes at the center of the image.

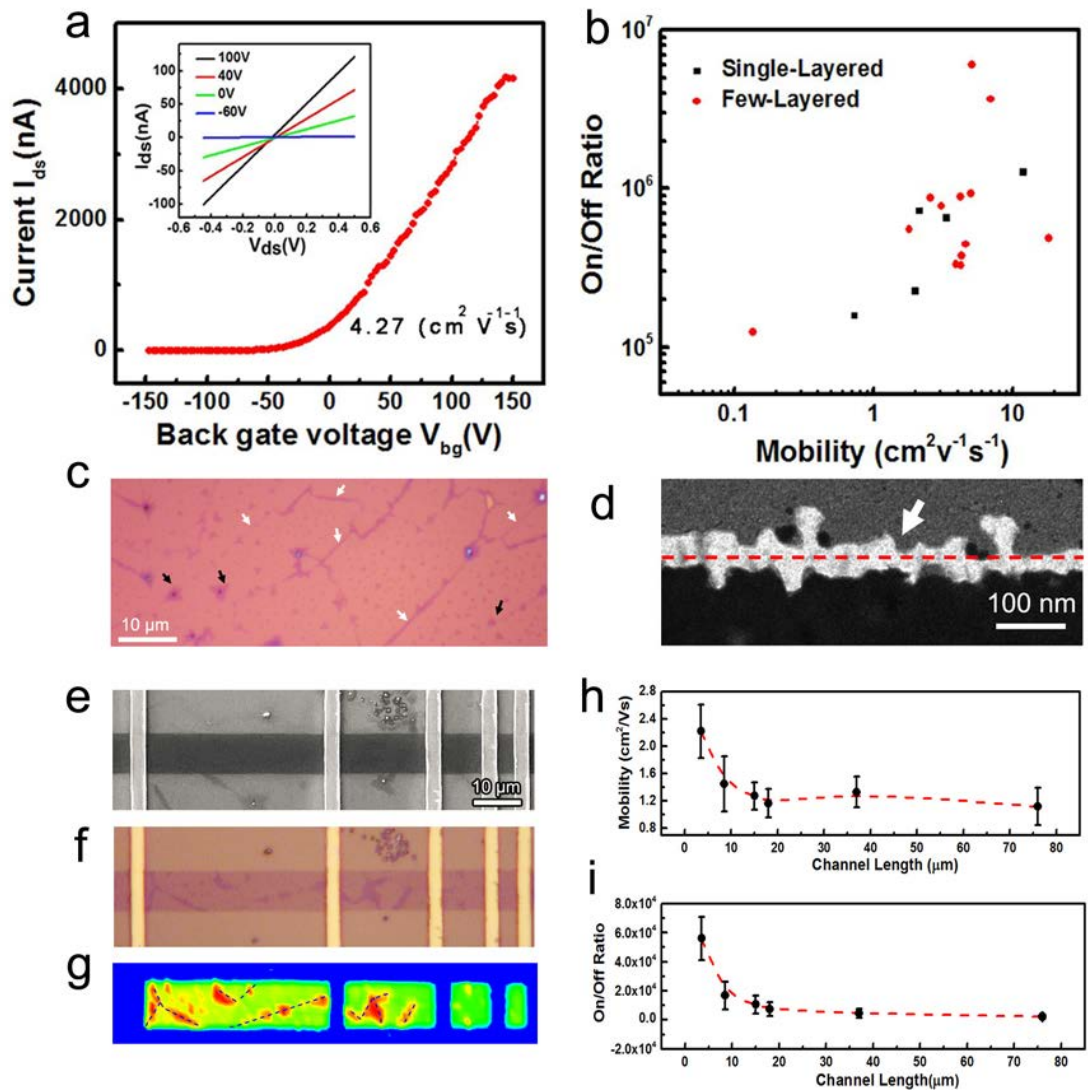

**Figure 5| Electrical properties and the effect of grain boundaries. a,** Room-temperature characteristics of the FET devices with 5 V applied bias voltage. Inset: Drain current-voltage curves acquired for back gating voltages of -60, 0, 40 and 100 V representing a good ohmic contact. **b,** A summary of the mobility and ON/OFF ratio measurements on MoS$_2$ FET devices, demonstrating a mobility range of 0.2 - 18 cm$^2$V$^{-1}$s$^{-1}$ and a maximum ON/OFF current ratio of 6×10$^6$. **c,** The optical image of the MoS$_2$ film showing the presence of a second MoS$_2$ layer grown along the grain boundaries, indicated by the white arrows. **d,** DF TEM image showing the grain boundary covered by a second layer, indicated by the arrow. **e** and **f,** The SEM and optical image of the device designed for the channel length dependency measurements. **g,** Corresponding Raman mapping of the same locations showing the distributions of the grain boundaries in the film, indicated by the dashed line. **h** and **i,** Channel length dependency of the charge carrier mobility and on-off ratio of the FET devices.

**Supplementary Materials:**

**Vapor Phase Growth and Grain Boundary Structure of Molybdenum Disulfide Atomic Layers**

**$MoO_3$ ribbon preparation**

As the precursor to $MoS_2$, $MoO_3$ films were prepared by the filtration or dispersion of their highly crystalline nanoribbons, which were produced hydrothermally through a process widely used for synthesis of this type of single-crystalline material [1]. Sodium molybdate ($Na_2MoO_4$) or heptamolybdate tetrahydrate $(NH_4)_6Mo_7O_{24} \cdot 4H_2O$ were used to synthesize $MoO_3$ nanoribbons. We dissolved 1.2 grams of these agents in nitric acid and transferred the solution to a Teflon container, which was heated at $170^oC$ for 1-2 hours. These $MoO_3$ nanoribbons have a high aspect ratio, roughly 20 microns in length, 1-2 microns in width, and a thickness in the range of 10-40 nanometers (Fig. S1a).

**CVD growth process**

The ribbons were then filtered or dispersed and large-area films were formed, cut into pieces, and placed on silicon substrates. This $MoO_3$-covered silicon substrate, along with several bare substrates designated for the growth of $MoS_2$ were placed close to each other at the center of the furnace vented with nitrogen at 200 sccm (Figs. S1b & S1c). At the opening of the furnace, a container with 0.8 - 1.2 grams of sublimated sulfur was placed at a location reaching an approximate maximum temperature of $600^oC$. The center of the furnace was gradually heated from room temperature to $550^oC$ in 30min at a rate of approximately $20^oC/min$. As the temperature approached $550^oC$, the sulfur slowly evaporated; the chamber was then heated to $850^oC$ at a slower pace of $\sim 5^oC/min$. After 10-15 minutes at this temperature, the furnace was naturally cooled back to room temperature.

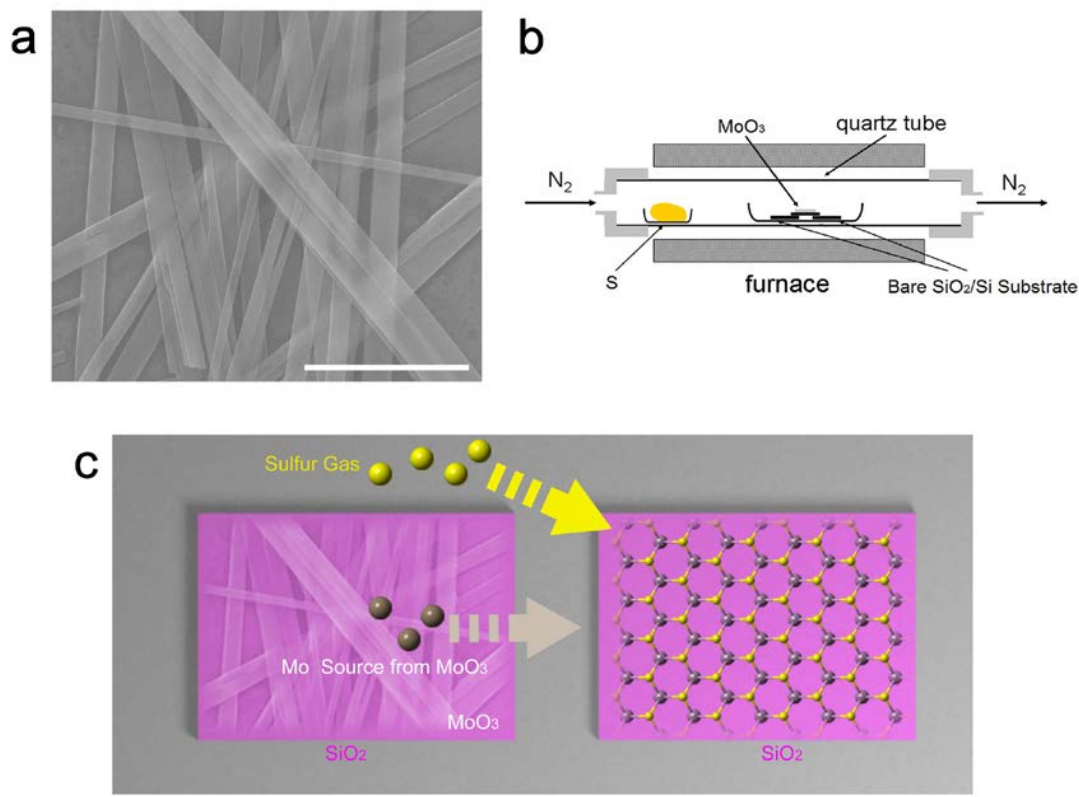

**Figure. S1|** Precursors and CVD setup. (a) Scanning electron microscopy image of $MoO_3$ nanoribbons prepared through a standard hydrothermal method and used as precursors in varied densities or condensed films. These ribbons provide a robust method for dispersion, allowing for the examination of the different stages involved in the growth process. The scale bar is 4 μm. (b) The configuration used in our experiments for $MoS_2$ film preparation. In this configuration, the $MoO_3$ precursors are dispersed on $SiO_2$/Si substrates and placed in the center of the quartz container, and several bare or patterned clean substrates, designated for $MoS_2$ growth, are positioned on the sides. The sublimated sulfur is placed close to the opening of the furnace, where it evaporates as the temperature at the center of the furnace approaches 550°C. Nitrogen is continuously streamed through the chamber with a flow rate of 200 sccm. (c) Schematic of the CVD growth of $MoS_2$ highlighting the vapor phase reaction between $MoO_{3-x}$ species and sulfur and the deposition of $MoS_2$ on a designated substrate.

**Proposed MoS$_2$ growth dynamics, effect of sulfur concentration and pressure on the growth process**

Although the full extent of the reaction between sulfur and MoO$_3$ has not been explored, some information can be extrapolated from the extensively-studied interaction between MoO$_3$ and H$_2$S [2, 3]. It is known that this reaction includes a transition to the MoO$_{(3-x)}$ species and the subsequent formation of oxisulfides [2]. At a suitable temperature range (200 - 400°C) and given sufficient reaction time, this reaction proceeds, and a complete conversion to MoS$_2$ occurs [2, 3]. It is postulated that similar stepwise transitions in vapor phase are involved in our experiments, in which H$_2$S is substituted with sulfur as reactant. When the intermediate oxides form, they diffuse across the bare substrates and, as a result of sulfurization, initially form the triangular MoS$_2$ domains and further grow into continuous films (Fig S1c). The temperature dependent studies reveal the ideal temperature for the high-quality formation of these triangular flakes is in the range of 800 - 850°C, in agreement with the previously reported experiments [3]. The sulfur concentration in this process at a given time of the experiment depends on the position of its container relative to the center of the furnace, the outgoing flow of gases from the chamber, and the initial loading of the sulfur. The experiments demonstrate that, in the case of an insufficient sulfur supply, oxisulfide rectangular domains nucleate and grow, instead of MoS$_2$ triangles and films (Fig S2a & b). As the sulfur concentration is increased to 0.5-0.7 grams, hexagonal MoS$_2$ domains nucleate, and at >1 grams of sulfur, triangular domains dominate (Fig S2c & d).

Another important parameter interrelated to the sulfur concentration is the pressure. As the sulfur evaporates the pressure increases. The out going flow can be adjusted to control the maximum pressure reached in the experiment, which also controls the amount of sulfur in the chamber. The maximum pressure is used in the following discussion as a measure for the pressure dependency of the MoS$_2$ growth. The measured pressure is the difference between the reference atmospheric pressure and the maximum pressure reached in these experiments. It is evident that both single-crystalline growth and density of nucleation are affected by the pressure. As the pressure increases less MoO$_3$ is evaporated, and its simultaneous solid state sulfurization slows down and ultimately stops the process.

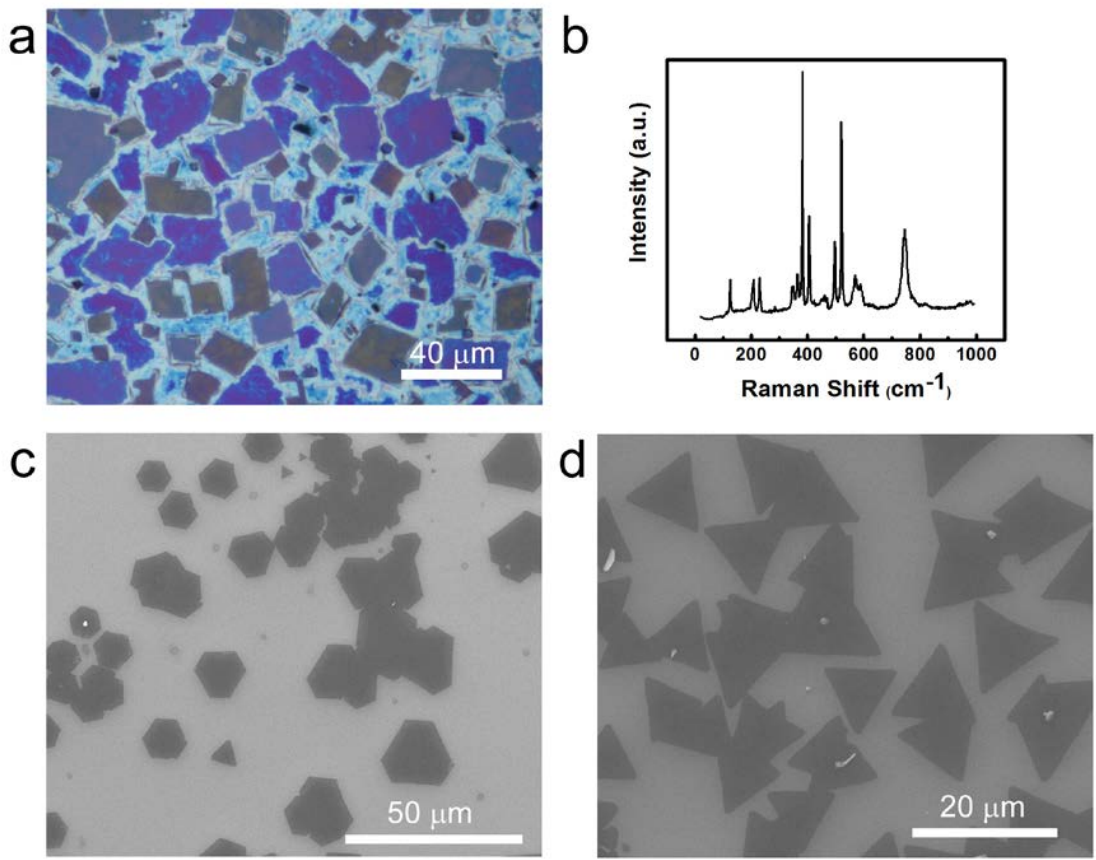

**Figure. S2|** Effect of sulfur concentration on the growth process. (a) At low sulfur concentrations, < 0.5 grams, rectangular domains are nucleated. (b) The Raman signature of these domains corresponds to oxisulfide ($MoOS_2$). The orthorhombic crystal structure of $MoO_3$ justifies the rectangular shape of these domains. Further sulfurization of the rectangular domains at temperatures of 800-1000°C reveals a significant solid phase stability of the material. The results are in agreement with our hypothesis that the triangular $MoS_2$ domains are nucleated as the vapor phase sulfur reduction of $MoO_3$ is completed on the surface of the substrate. (c) Nucleation of $MoS_2$ begins at sulfur concentrations in the range of 0.5-0.7 grams; however, the domains construct a hexagonal geometry. (d) At higher sulfur concentrations, >0.7 grams, triangular domain nucleation dominates the process.

In Fig. S3, the changes in the growth process is demonstrated by showing the most commonly observed features. At close to atmospheric pressures (0.5-1KPa) the supply of sulfur is low but sufficient MoO$_3$ is evaporated and typically oxisulfides or hexagonal isolated islands form (Fig S2). Additionally at the same pressures granular and thick films may grow (Fig S3a). Increasing the pressure to 1-4 KPa, the growth of small triangles initiates (Fig S3b) and it extends to large area films in the pressures range of 4-10K (Fig S3c). At these pressures both supplies of MoO$_3$ and sulfur are sufficient for large area growth and coalescence of triangular domains. At slightly higher pressures, 10-40KPa sulfur is plentiful but lower evaporation of MoO$_3$ slows down its supply and isolated but very large, 50-80 μm, triangular domains form (Fig 3d). At higher pressures, 40-80KPa, these isolated triangles shrink in size and their morphology begins to change (Fig 3e). At the maximum pressures that the CVD chamber can withhold, 80-120KPa, isolated star shaped islands grow (Fig 3f). In conclusion the sulfur concentration is the limiting factor that determines the morphology and the size of the single crystalline domains and MoO$_3$ controls the density of nucleation and large-area growth of MoS$_2$.

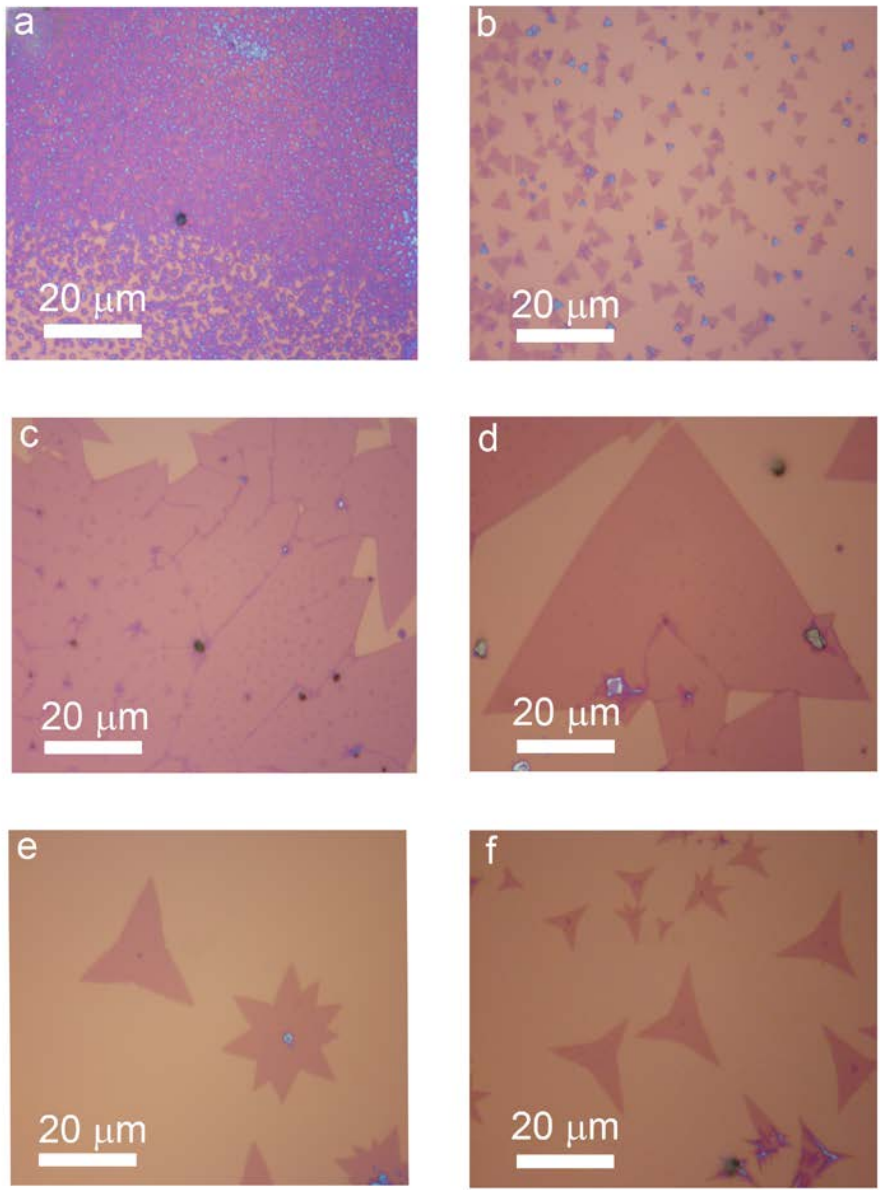

**Figure 3|** Effect of pressure on the growth processes. (a) Granular films at 0.5-1KPa. (b) High density nucleation of triangular domains at 1-4 KPa. (c) Formation of large area films at 4-10KPa. (d) Growth of large area triangular domains at 10-40KPa. (e) Formation of isolated triangular domains with modified shapes at 40-80KPa. (f) Star shaped domains at 80-120KPa

**Controlled Nucleation and Growth of MoS$_2$**

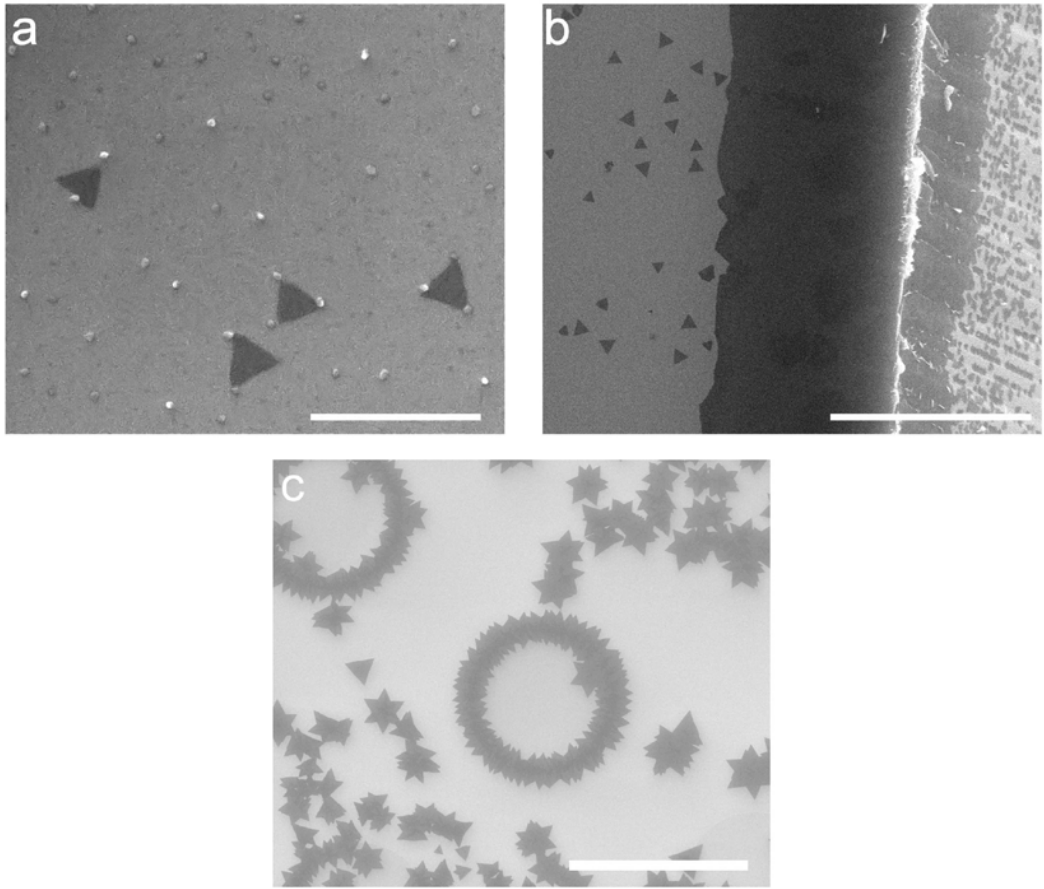

**Figure. S4|** Common features observed in the direct synthesis of MoS$_2$ on substrates. (a) SEM image showing the nucleation sites formed across the bare substrate in a random manner, leading to the formation of triangular MoS$_2$. (b) SEM image demonstrating the affinity of nucleation to the edges and rough surfaces. As evident from this image, a higher density of nucleation is frequently observed on the cross-sectional surface and at substrate edges. (c) Induced nucleation of MoS$_2$ domains near artificially-made circular edges on SiO$_2$. Some of the nucleated domains diffuse from these edges to flat surfaces. The scale bars are 1, 10, and 50 μm for (a), (b), and (c), respectively.

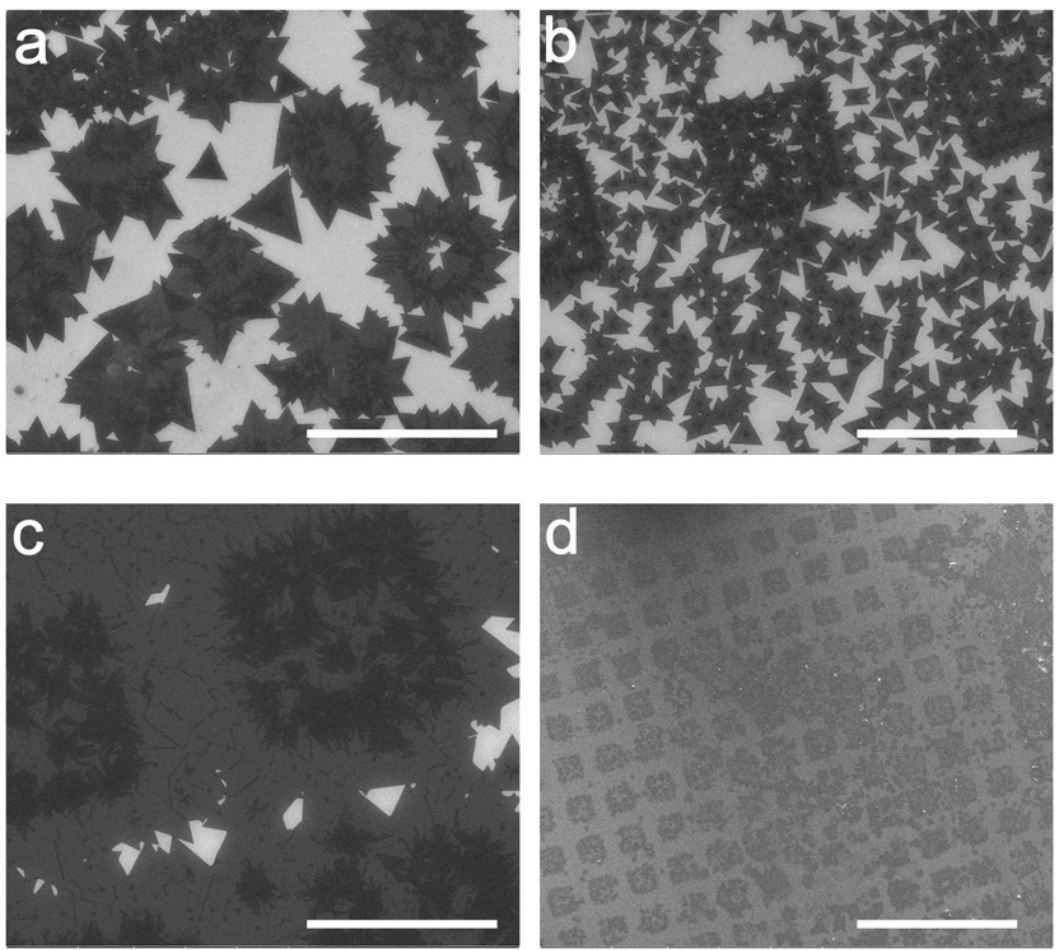

**Figure. S5|** Nucleation controlled growth process. (a-d) SEM images showing the nucleation controlled growth process through the patterning of silicon substrates. These patterns are discernable by the obvious contrast resulting from dense nucleation. Film growth is initiated by the nucleation and growth of triangular single-crystalline $MoS_2$. A sufficient supply of $MoO_3$ results in further growth and coalescence of such triangles, and ultimately, continuous large-area films with high level of crystallinity form. The scale bars in (a-d) are 100, 50, 50, and 400 μm, respectively.

The sulfur concentration and pressure controls provide the means to maximize the size of the single-crystalline domains, before they merge. Our experiments at lower $MoO_3$ concentrations, which allows for resolving the grains before they coalesce, show that a maximum triangular domain sizes in the order of 20-25 μm should be expected. This is in agreement with the results described for the pressure dependency of the $MoS_2$ growth, which illustrates the correlation between the maximum domain sizes and large area growth of the MoS$_2$ films. The existence of a maximum achievable grain size imposes restrictions on the pattern base strategy and control of crystallinity in the large area films. In Fig. S6a, where 20 μm$^2$ rectangular patterns with the spacing of 20 μm are used, although most triangular domains have not reached their maximum size (~20 μm), they have already made contact with triangles nucleated from other sites and, as more precursor is provided, over-layers grow (Fig. S6b). However, for slightly larger patterns, 40 μm$^2$ rectangles with spacing of 40 μm, many triangles in the maximum size range are observed and they meet other domains at relatively larger sizes (Figs. S6c and d). Our experiments show that further pattern size increase results in insufficient nucleation and deters the large area growth. These results exemplify a general control mechanism in the pattern base growth for control of film size and crystal quality.

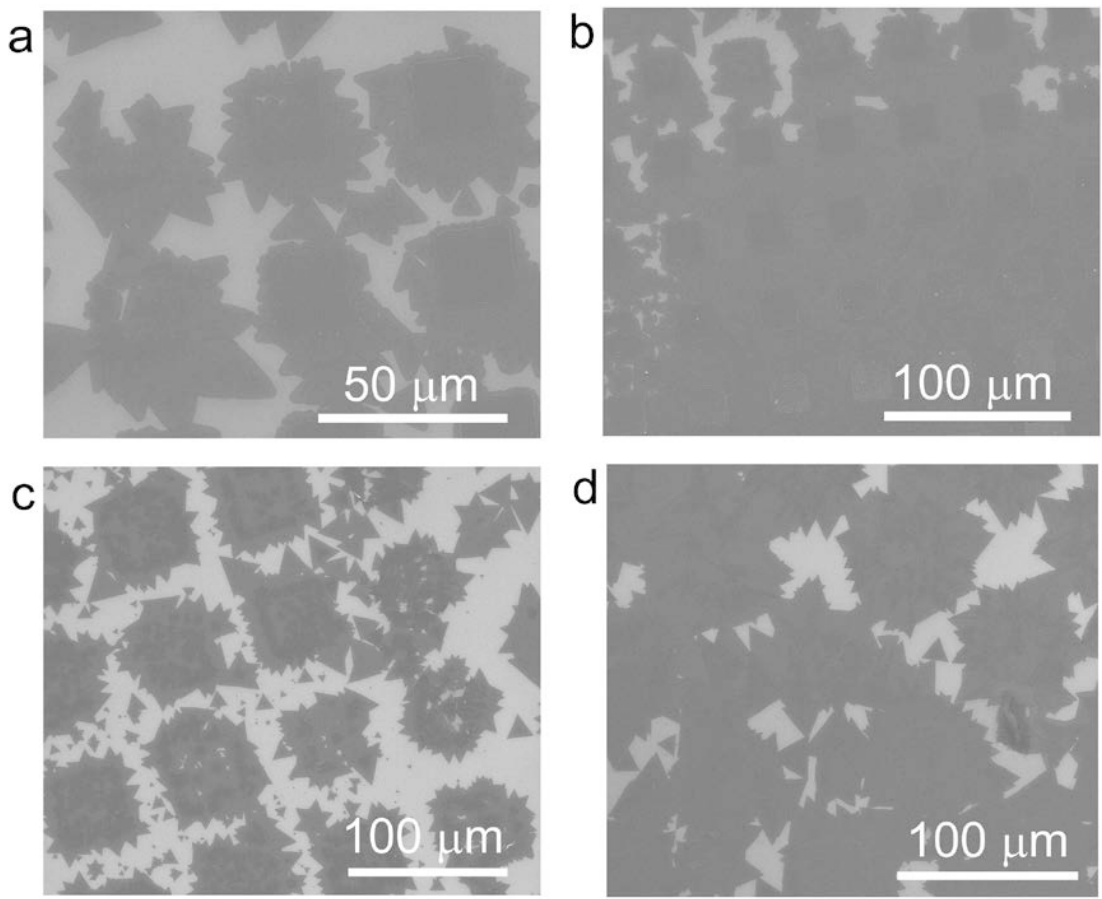

**Figure. S6|** Method for control of crystallinity in MoS$_2$ films through pattern-size variation. (a) MoS$_2$ synthesis at low densities of MoO$_3$ on 20 μm × 20 μm rectangular patterns located 20 μm apart shows that triangles with sizes ~10 μm have already made contact with triangles nucleated from other sites, suggesting an average grain size in the order of <10 μm for the films made on this type of patterns. (b) A further supply of MoO$_3$ results in the formation of films with high possibility of over-layer nucleation and over-

layer growth. (c) MoS$_2$ synthesis using 40 μm × 40 μm rectangular patterns located 40 μm apart at low source supply. Triangular domains within the size range of 20 μm are commonly observed. (d) In these patterns, sufficient source supply allows domains of relatively larger sizes to merge into more continuous and high-crystalline films.

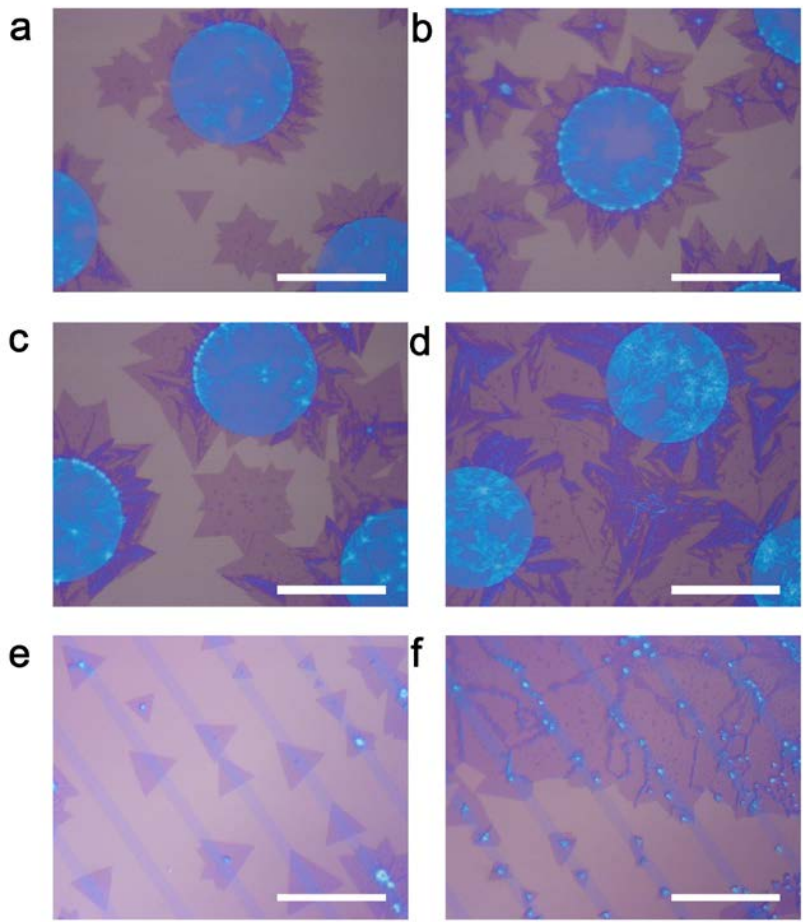

**Figure. S7|** Variation in patterns for the controlled nucleation and growth process. (**A-D**) The application of circular patterns demonstrates a similar nucleation and growth process. However, since growth in multiple directions from the initial nucleation sites at the edge of the circles is possible, overlapped and multi-layer growth is more prominent. (**E-F**) Growth using patterns with long rectangular bar where the nucleation is clearly affiliated to these bars; however, smaller surface area results in less surface nucleation, and therefore results in slower formation of the films. An incomplete and scattered formation of ~100 μm films results from the diffusion-limited nature of the synthetic process. The scale bars in all images are 40 μm.

**Optical examination, Raman measurements, XPS and EELS**

The optical images of the films and the triangular domains provide a qualitative measure of the samples' thickness and uniformity (Fig. S8a). We further evaluate their thickness and surface roughness using Raman spectroscopy. It has been shown that the two vibrational modes of MoS$_2$ tend to approach each other down to ~18 – 20 cm$^{-1}$ in single-layered samples, providing a robust thickness characterization tool for samples up to and including 4 layers [4]. The measured differences between the characteristic Raman peak positions, E$_{2g}^1$ and A$_{1g}$ – acquired from areas presented in Fig. S8a – are 19.5 and 22.2 cm$^{-1}$, corresponding to the thicknesses of one and two layers (Fig. S8b). Our thickness analyses across these samples indicate that these films are mostly single-layered, with occasional two- or few-layered regions. Raman intensity mapping has been proposed as a method for the investigation of grain boundaries in graphene [5]. These studies showed that the D-band intensity in graphene could be utilized to detect nucleation sites and grain boundaries. In the present study, the Raman intensity mapping for characteristic band positions of single-layered samples, E$_{2g}^1$ (385cm$^{-1}$) and A$_{1g}$ (405cm$^{-1}$), from the merged triangles are collected. These results reiterate the thickness uniformity of these samples (Fig. S8c and d). In these maps, the nucleation sites are clearly discernable; however, no clear indications of changes in the vibrational properties are observed at likely grain boundaries. XPS measurements at every stage of the growth process, from the triangular domains to continuous films, provide strong evidence for the high-quality formation of MoS$_2$, as presented in Figs S9a and b. The information acquired from the position and intensity of the Mo 3$d$ and S 2$p$ bands on dispersed triangles, discontinuous films, and completed films demonstrate that the bonds in MoS$_2$ samples at every stage are fully formed. In these samples, the single molybdenum doublet Mo 3$d_{3/2}$ and Mo 3$d_{5/2}$ appear at binding energies of 233.9 eV and 230.9 eV, and the sulfur doublets S 2$p_{1/2}$ and S 2$p_{3/2}$ at 162.9 eV and 161.8 eV, respectively. These measurements match the characteristic band positions of fully-transformed $MoS_2$ during the sulfurization of $MoO_3$, validating its complete conversion during the growth process [3]. Moreover, electron energy-loss spectroscopy (EELS) obtained from clean regions in the $MoS_2$ domains shows only the characteristic peaks of Mo (M-edge) and S (L-edge) without impurities (Fig. S9c), which further confirms these findings.

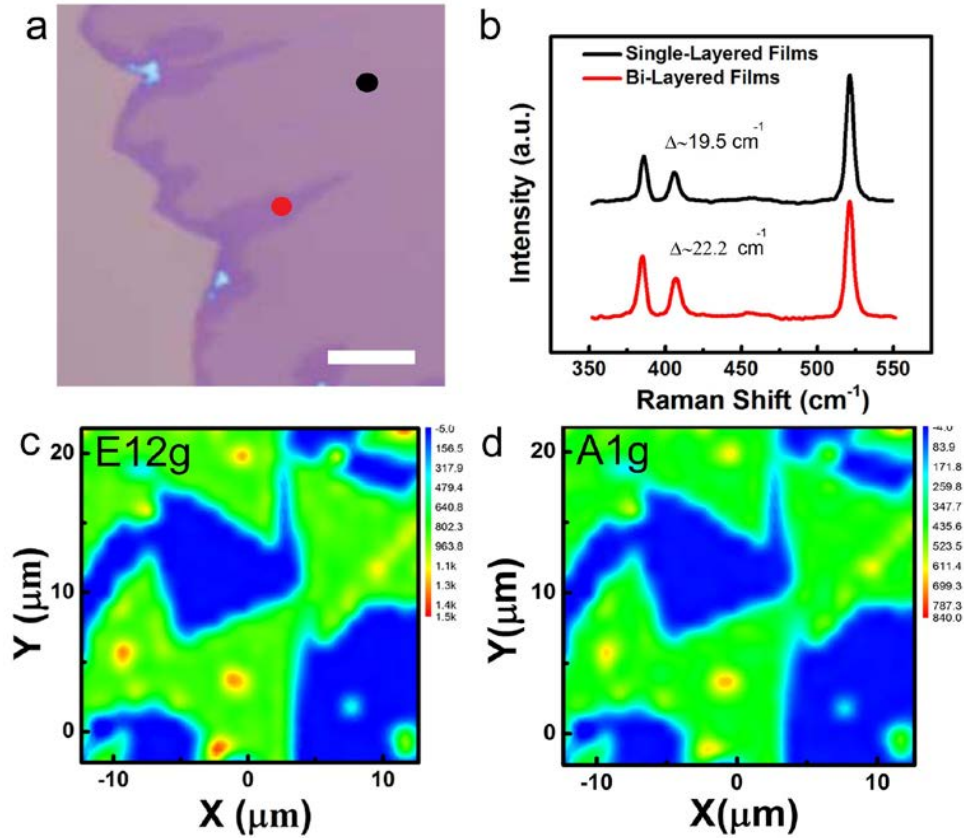

Figure. S8| **Thickness and chemical characterization of as-synthesized $MoS_2$ samples. (a) Optical image of $MoS_2$ films with optical contrast resembling single layered samples; the scale bar is 40 μm. (b) Raman measurements from the two points marked on (a)** by red and black filled circles **demonstrating the thickness variations in these samples.** (c) The in-plane $E_{12g}$ and (d) out-of-plane $A_{1g}$ modes of vibration positioned at 385 and 405 cm$^{-1}$, respectively, corresponding to single-layered $MoS_2$. These maps demonstrate the thickness consistency in these triangles; it is evident that these triangles consist of nucleation centers made of $MoS_2$ particles with larger thickness. Grain boundaries are not clearly detected based on the Raman measurements.

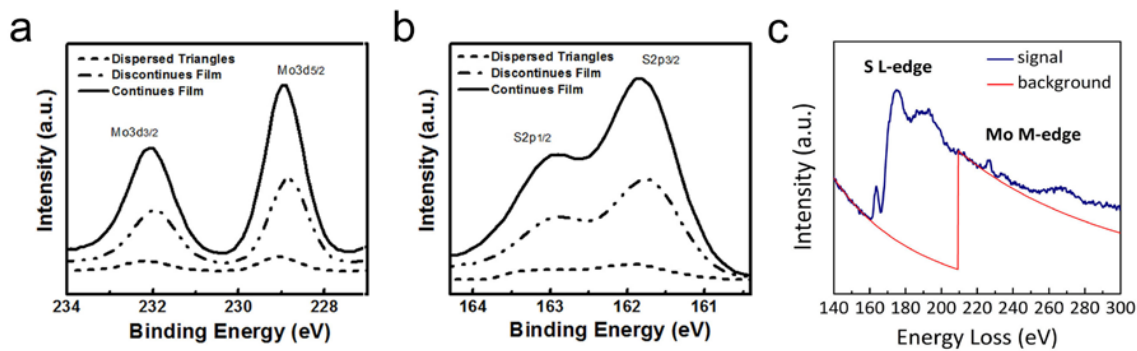

Figure. S9| **Chemical analysis of MoS$_2$ single triangular domains and films. (a-b) The XPS measurements revealing the complete formation of MoS$_2$ at every stage of the growth process,** *i.e.* **from the nucleation of small triangles to continuous films. (c)** EELS spectrum acquired from the defect-free MoS$_2$ monolayer film, showing the S L-edge and Mo M-edge.

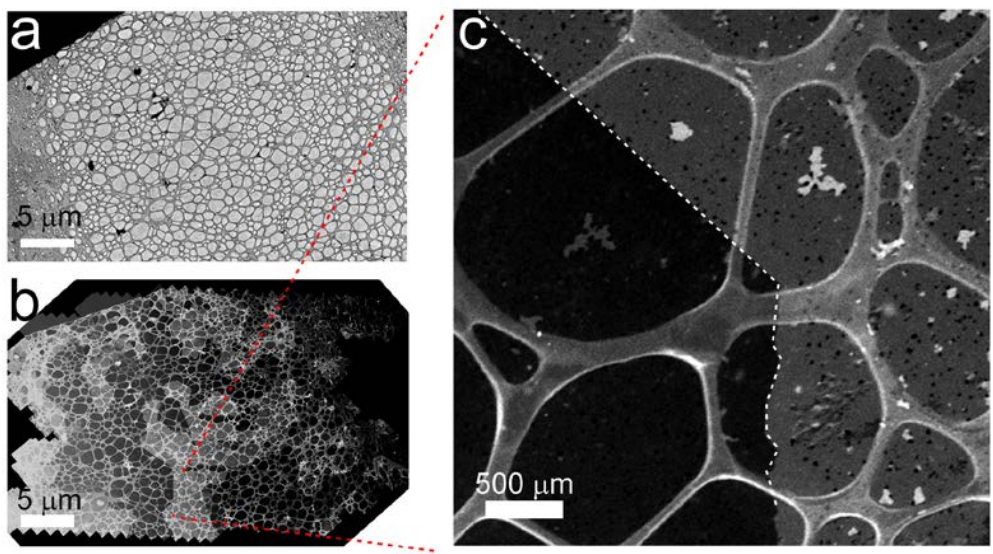

**Figure. S10|** BF and DF TEM images of MoS$_2$ films. (a) BF TEM image of the MoS$_2$ film. (b) DF TEM image of the same region showing a large MoS$_2$ grain across the whole film. The DF image shown in (b) was obtained by stitching 507 different DF images, acquired with a spacing of 3 µm under the same microscope conditions. (c) An individual DF image aquired from the square area shown in (b).

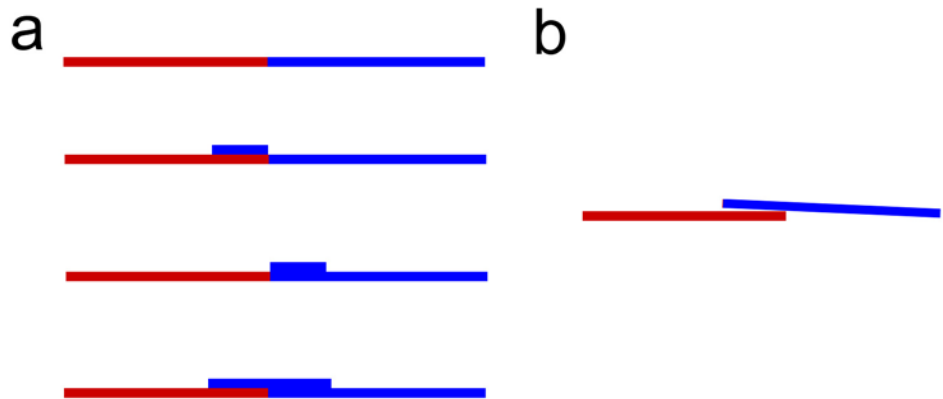

**Figure. S11|** Two modes of boundary formation. Grains with different orientations are indicated by red and blue lines. (a) Tradition grain boundaries involve the formation of chemical bonds between the two in-plane grains at the boundaries. The growth process can further extend through the nucleation of a new layer with crystal orientation similar or different from one of the original in-plane grains. (b) Overlapped grain junctions are a mode of boundary formation, in which no chemical bonds between the two in-plane grains form and the two grains may grow on top of each other. This mode can be distinguished from the traditional grain boundaries in (a) through high-resolution electron microscopy imaging, where a discontinuity in the lattice structure of conventional grain boundaries is noticeable (see Fig. 4 in the main text).

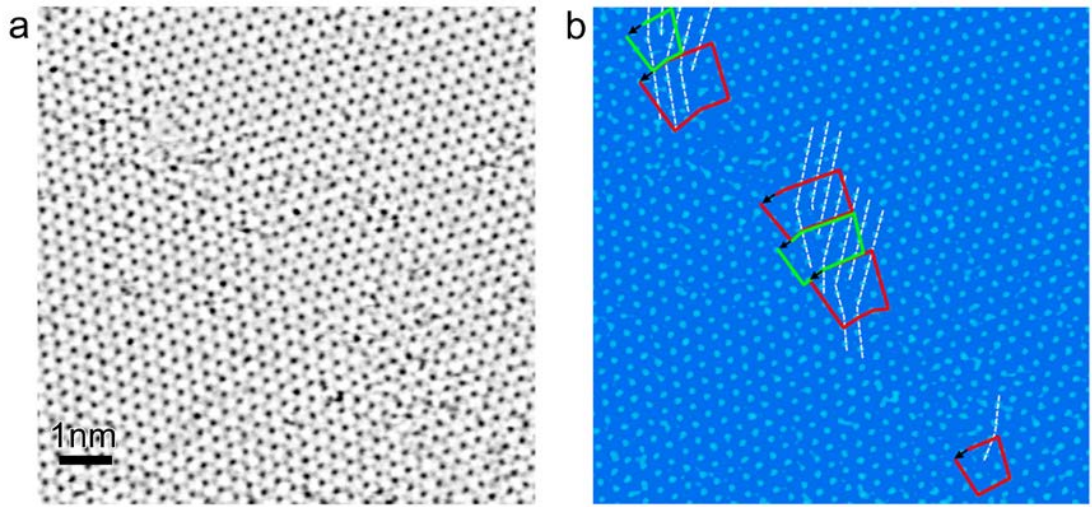

**Figure. S12**| Burgers vector analysis of the MoS$_2$ grain boundaries. (a) A negative-contrast STEM-ADF image of a 21° MoS$_2$ grain boundary. (b) False-color image of (a), with Mo atoms highlighted. Burgers vectors are shown as black arrows, and are calculated by drawing Burgers circuits (shown as red and green lines) and by recording the vectors that would complete the circuits. Following the same notation for Burgers vector in graphene with hexagonal lattice, all dislocations have the shortest Burgers vector (1, 0) with the smallest energy, and with a magnitude proportional to its squared Burgers vector. The result strongly suggests that dislocations can be constructed by inserting or removing a semi-infinite stripe of atoms along the armchair direction in the Mo-oriented lattice

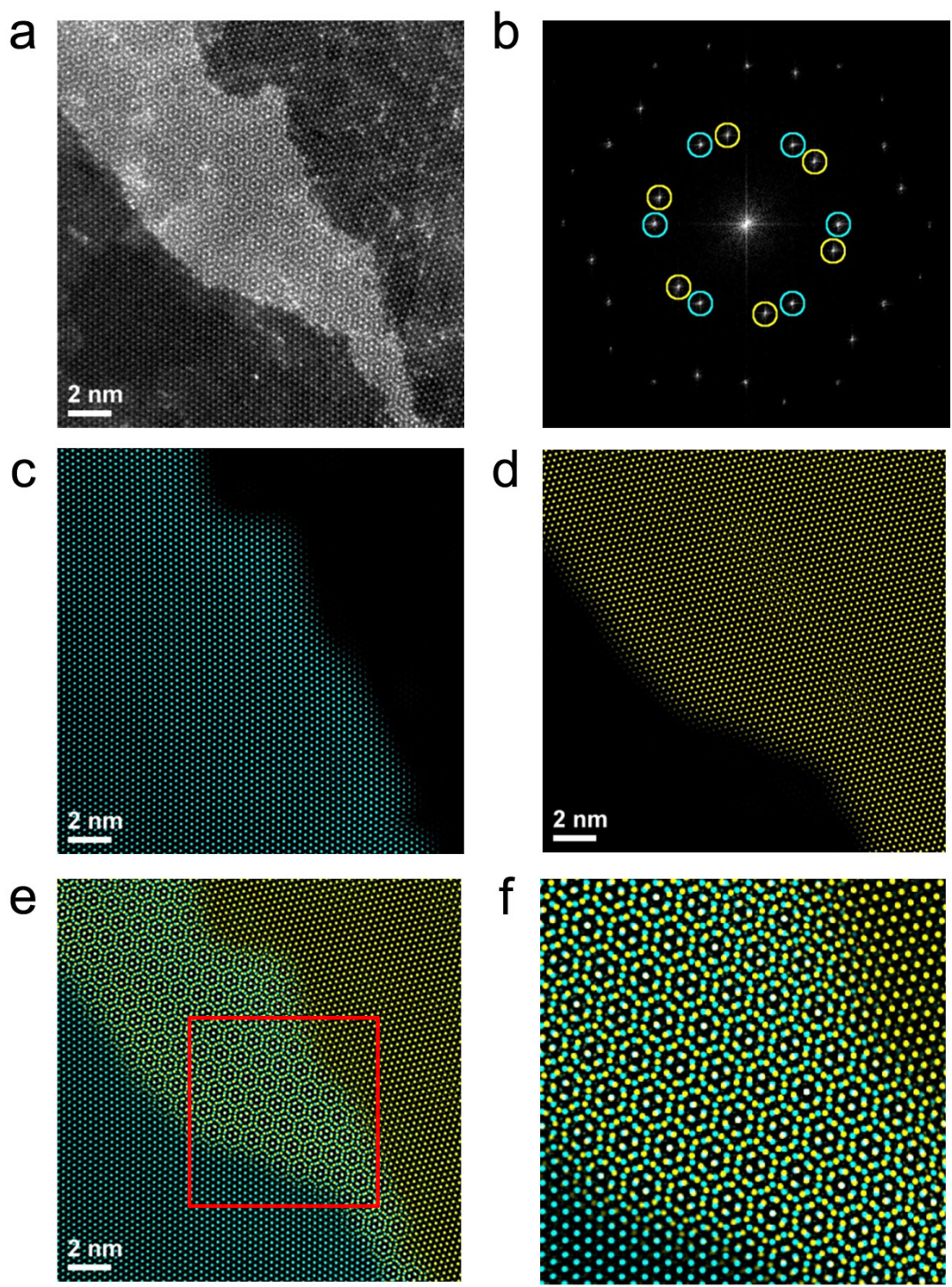

**Figure. S13| |** STEM-ADF imaging of an overlapped junction between two monolayer MoS$_2$ grains with 17º rotation. (a) Low-pass filtered ADF image. (b) FFT of the image showing two sets of MoS$_2$ diffraction spots with 17º rotation. (c) FFT-filtered image using one set of diffraction spots highlighted in cyan in (b). (d) FFT-filtered image using one set of diffraction spots highlighted in yellow in (b). Note that there is no

discontinuity in the lattice structure within each grain at the junction point, confirming that the two layers continue to grow on top of each other without forming in-plane chemical bonds. (e) False-color FFT-filtered image constructed from (c) and (d). (f) Magnified view of the highlighted region in (e) showing the distinct Moiré fringes generated from the overlapping of the two grains.

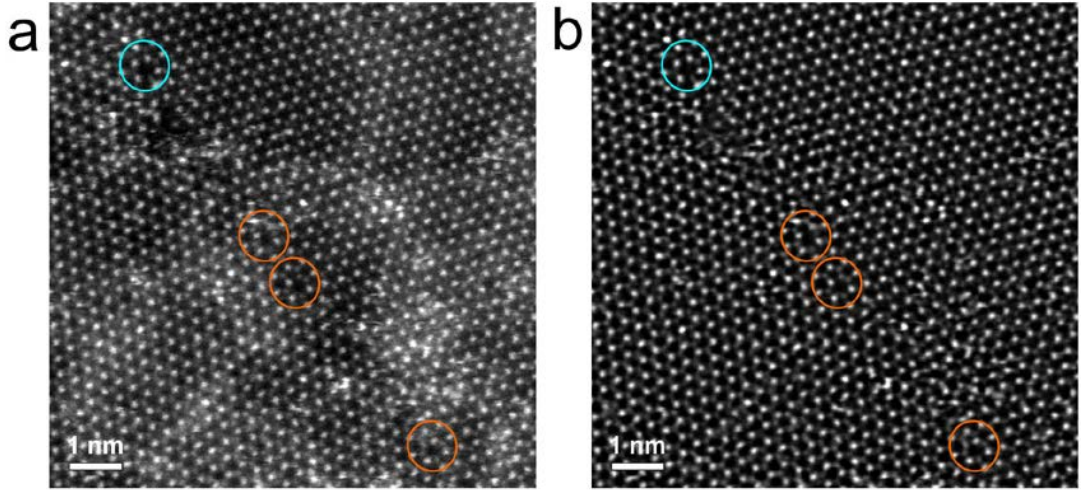

**Figure. S14|** STEM-ADF image of a grain boundary with 21° tilt angle. (a) Low-pass filtered image. (b) Image after low-pass filtering and partial filtering of the direct diffraction spot in the FFT, which aims to remove the contrast variation due to surface contamination. The light blue circles highlight the positions of Mo-oriented 5|7 dislocation cores at the grain boundary, and the orange circles highlight Mo-oriented 5|7 dislocation cores with an $S_2$ substitution.

**Density functional calculations**

The total energy first-principles calculations are performed using density functional theory (DFT) within the local density approximation (LDA) and the projector-augmented wave (PAW) method [6, 7], as implemented in Vienna *Ab-initio* Simulation Package (VASP) [8, 9]. The grain boundaries (GBs) are modeled with similar tilt angle (~ 21°) as those in the experiment. The periodic models incorporate oppositely-aligned Mo-oriented 5|7 in a large supercell. The in-plane lattice parameters are roughly 34 Å and 11 Å in the perpendicular direction and along the GBs, respectively. Between the layers, a vacuum layer of 12 Å is introduced and in all structures, a plane-wave-based total energy minimization scheme [10] – 1×3×1 Monkhorst-Pack k-point mesh centered at the Γ- point with an energy cutoff of 280 eV – is applied until the force on each atom is less than 0.01 eV/Å.

Formation energies for $S_2$ substitution are defined by:

$$E_f = E_{S2-sub} - E_{Mo-5|7} - \sum_i \Delta n_i \mu_i$$

Here, $E_{S2-sub}$ and $E_{Mo-5|7}$ are the total energies of $S_2$-substituted and un-substituted Mo-oriented 5|7s within similarly-sized supercells, respectively; $\Delta n_i$ and $\mu_i$ are the change in the number of atoms and the chemical potentials of species *i*, for Mo or S. In thermodynamic equilibrium with the $MoS_2$, $\mu_{Mo}$ and $\mu_S$ satisfy the equation $g_{MoS2} = \mu_{Mo} + \mu_S$, where $g_{MoS2}$ is the Gibbs free energy of an infinite $MoS_2$ sheet per unit molecule. The formation energies are calculated under the S-rich condition, taking chemical potential of the bulk-S as reference.

**Optical devices for opto-electrical characterization of $MoS_2$**

To demonstrate the optical properties and device performance of the $MoS_2$ films, photolithography was used to prepare electrodes on these samples for photocurrent measurements (Fig. S15b). We measured the voltage dependence of the devices' photoelectric response and calculated an ON/OFF ratio of ~10 and ~3 for single- and multi-layered devices, respectively (Fig. S15c and S15d). These results highlight a significant enhancement in the photocurrent response and photosensitivity of the single-layered, as compared to the multi-layered samples. We also measured the wavelength dependence of the photo-current for wavelengths ranging from 750 nm down to 200 nm (Fig. S15.e). The experiments reveal photocurrent properties similar to those observed in single-layered and bulk samples [11, 12]. As shown in Fig. S15e, the photocurrent signal drastically increases as the wavelength decreases to around 700 nm, corresponding to the absorption energy edge (~1.8eV) of single-layered $MoS_2$.

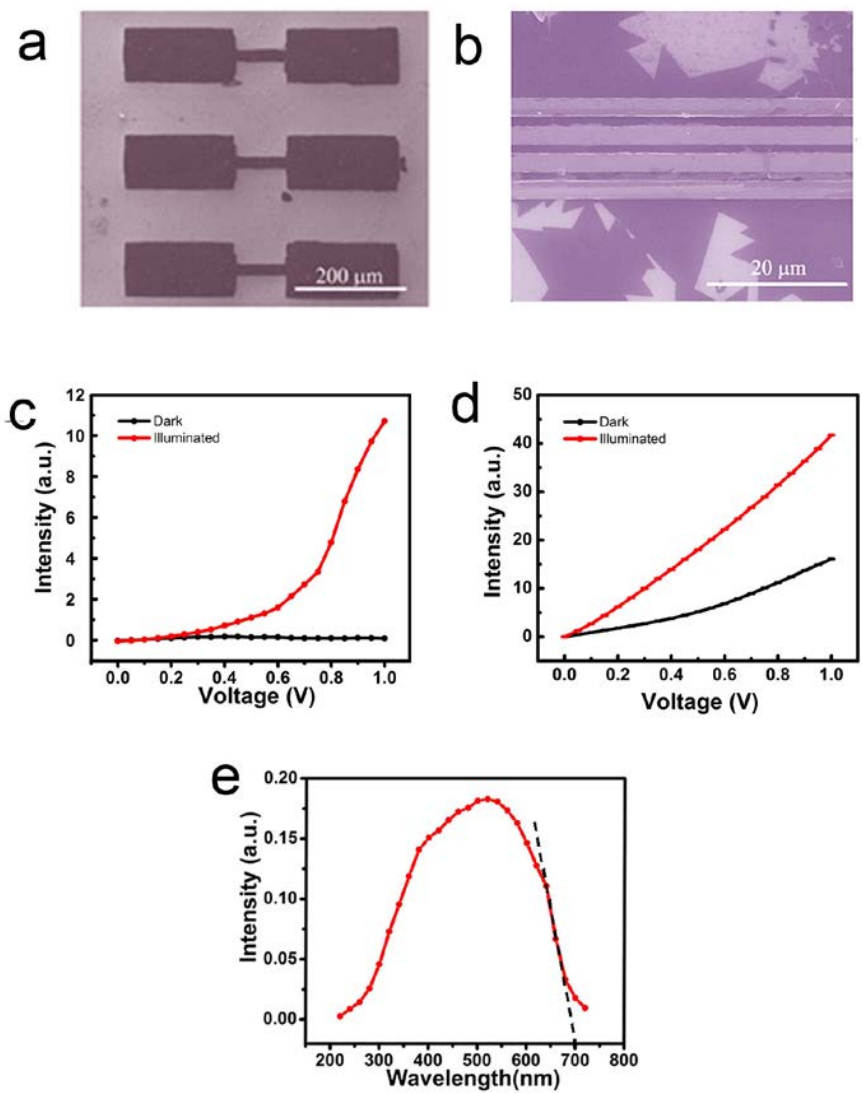

**Figure. S15|** Phototransistors and Photocurrent measurements. (a) FET device arrays prepared using lithography process. (b) Phototransistors for photocurrent characterization. At a laser wavelength of 405 nm, the photocurrent response of the $MoS_2$ films is measured for (c) single-layered flakes and (d) few-layered samples. For single-layered flakes, an ON/OFF current ratio of ~10 is demonstrated; for few-layered samples, this ratio was measured to be ~3. These results reveal a significant enhancement in the ON/OFF current ratio for the single-layered devices. (e) Wavelength dependency of the photocurrent in single-layered samples reveals the broad absorption band in this material. The onset of the photocurrent at ~700nm, as indicated by the dotted line, is associated to the bandgap energy of single-layered samples ~ 1.8 eV.

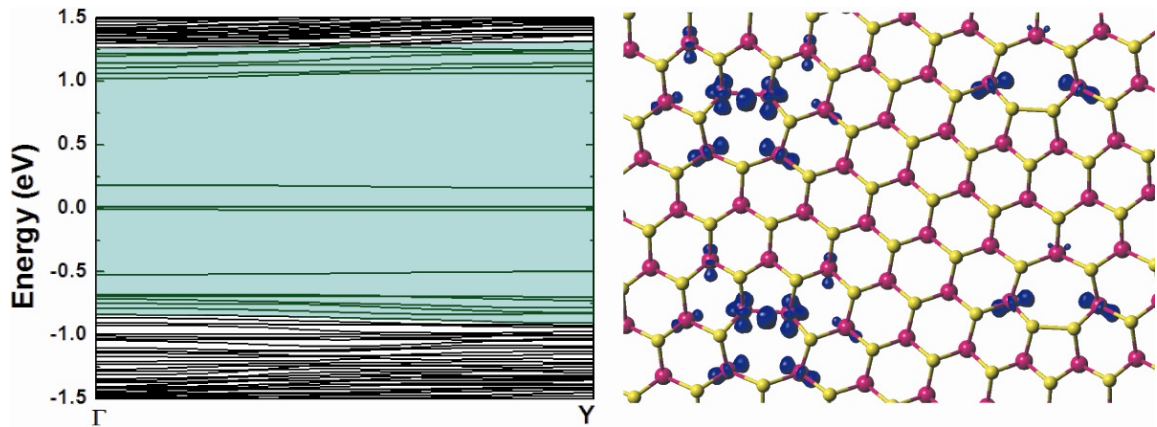

**Figure. S16|** Effect of grain boundaries on the electronic properties of $MoS_2$. In the left panel, the Fermi level is set to zero and the faded areas correspond to the localized energy bands mainly contributed by the grain boundaries. The electrons occupying these deep levels are highly localized and do not contribute to the conductivity, through thermal activation. Hence, these states serve as undesirable carrier sinks and degrade the device performance. Right panel schematically shows the partial charge density distribution corresponding to the energy range -0.5 to 0.5 (eV).

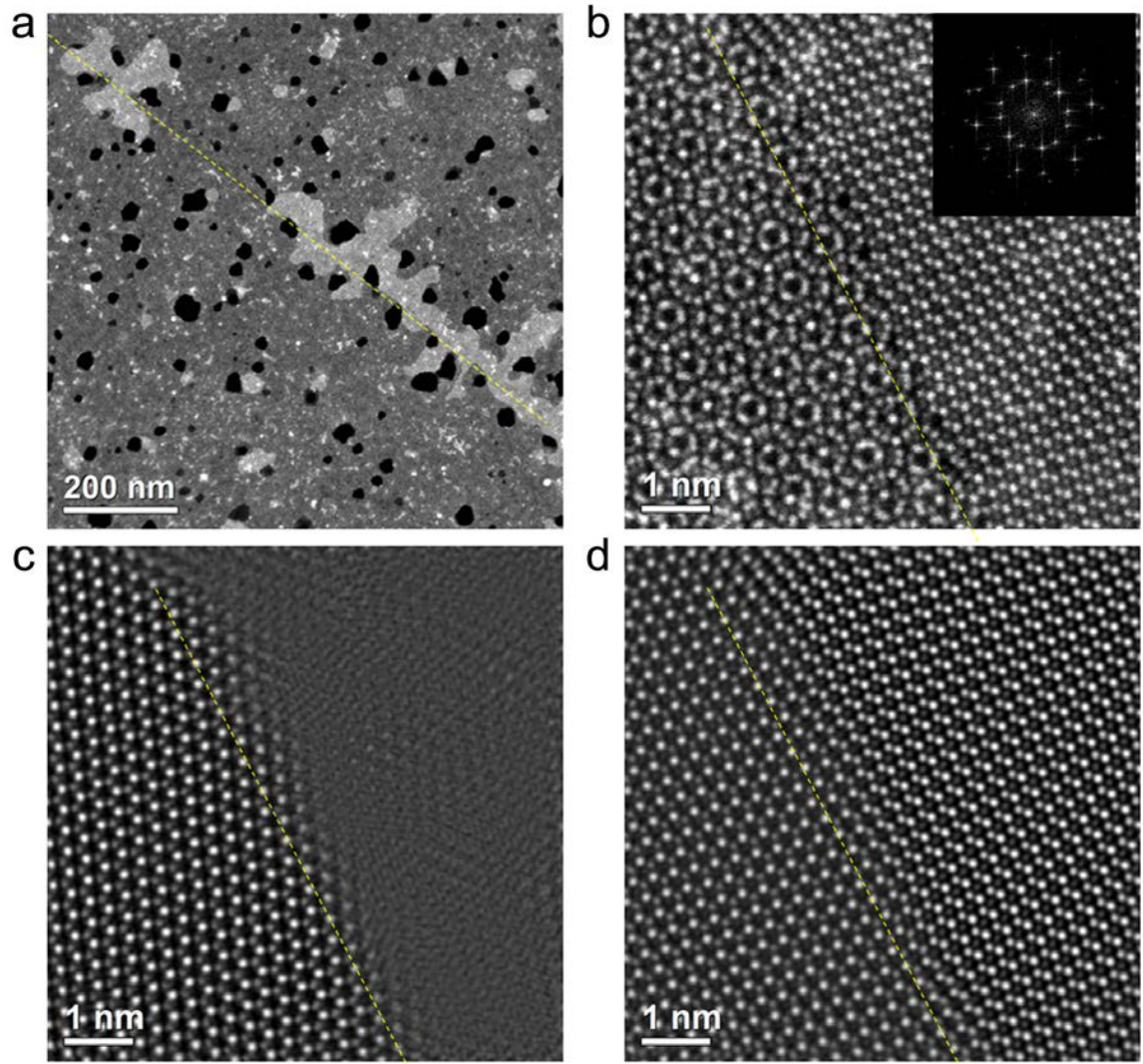

**Figure. S17 |** STEM-ADF images of the second MoS2 layer grown along the grain boundaries. (a) Low magnification ADF image showing the formation of second layer along the grain boundary (highlighted). (b) Atomic resolution ADF image showing the grain boundary covered by a second layer. (c) Fourier filtered image of (b) showing the position of one grain in the first layer. (d) Fourier filtered image of (b) showing the position of the other grain in the first layer and the second layer with the same orientation. The grain boundary is highlighted by the yellow dashed lines.

**Effect of line defects on the transport properties of MoS$_2$**

To further elucidate the role of grain boundaries in the transport properties of MoS$_2$, additional experiments were performed. We quantify the line defect density by measuring the total length of line defects – observable in the optical images – and divide it by the total area of the material (Fig. S18a). Fig. S18b and S18c demonstrates that the mobility and on/off ratio in devices with similar channel length decrease as a function of line defect density. It is also clear that the trends of these changes are independent of the applied bias and excludes the role of device contact resistance, which is significantly dependent on the applied bias. Similar measurements on a MoS$_2$ ribbon using a device with channel lengths ranging from 3.5 µm to 76 µm show that the channel length dependency described in the main text is predominantly caused by changes in density of line defects (Fig.S18d and e). Since larger channels have a higher chance of containing grain boundaries, a general channel length dependency in the mobility and on/off ratio is expected. However, line defect density is the governing reason for these changes; for small channel lengths containing a high density of line defects, one can observe the deviations from this general rule, explaining the observed large error bars seen in Fig.5h and i.

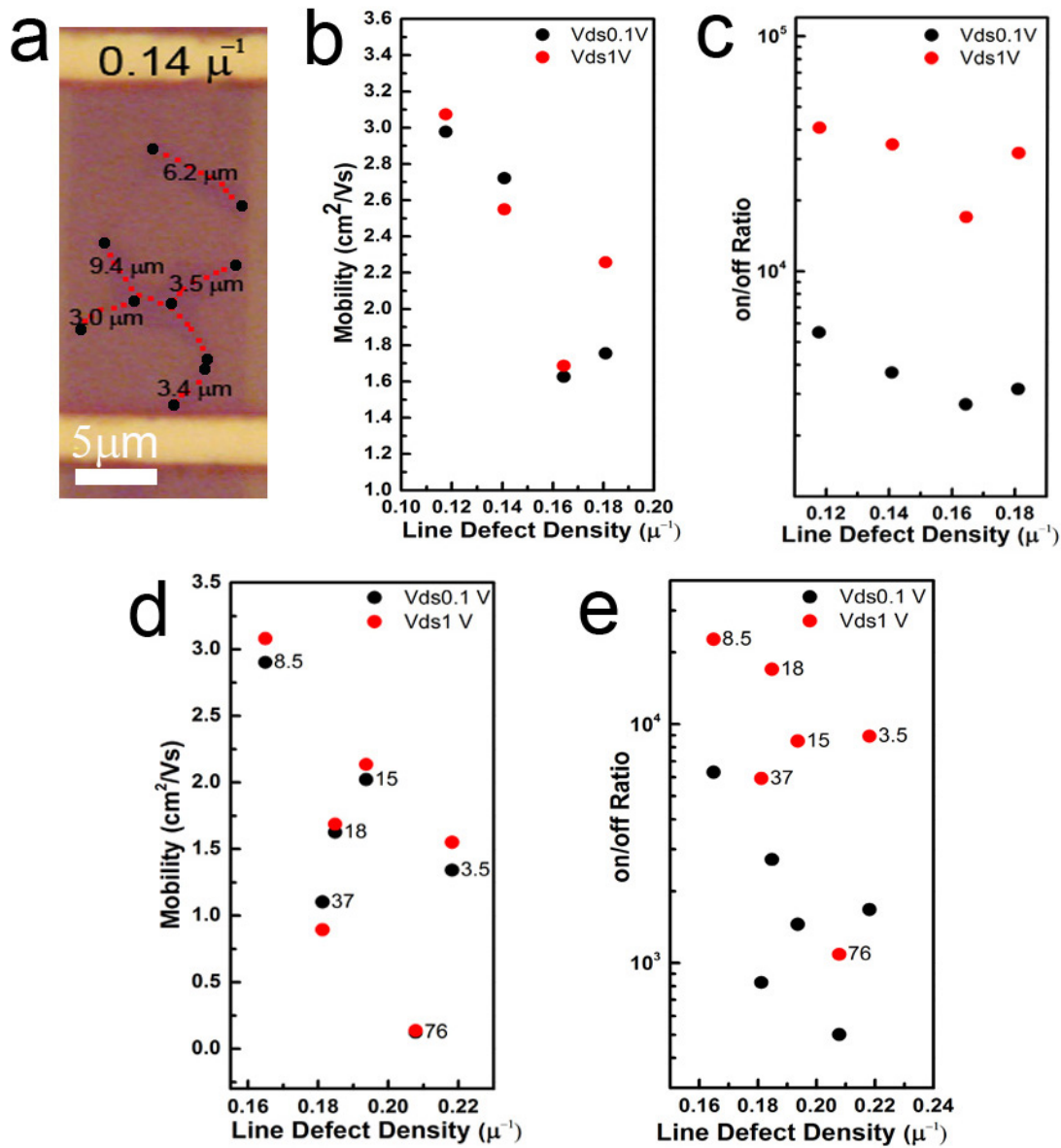

**Figure. S18** | Effect of grain boundaries on the transport properties of $MoS_2$. (a) The estimated density of line defects, $0.14\ \mu^{-1}$, in a typical $MoS_2$ ribbon. (b and c) Changes in the mobility and on-off ratio of devices with 18μm long channel and varied line defect density at $V_{ds}$= 0.1 and 1V. (d and e) Changes in the mobility and on/off ratio of a devices made on the same ribbon with different channel lengths as a function of line defect density at $V_{ds}$= 0.1 and 1V. The channel length in micrometers for each device is marked on the data points.